\pdfoutput=1
\documentclass[aip,pof, preprint]{revtex4-1}
\usepackage{calc}
\usepackage{array}
\usepackage{amsmath}
\usepackage{amsfonts}
\usepackage{amssymb}
\usepackage{amsthm}
\usepackage[dvips]{color}
\usepackage{graphicx}
\usepackage{epsfig}
\usepackage{subfigure}
\usepackage{algorithmic}
\usepackage{calc}
\usepackage{psfrag}
\usepackage{relsize}
\usepackage{color}
\usepackage{algorithmic}
\usepackage{algorithm}
\usepackage{float}

\floatstyle{ruled}
\newfloat{algo}{thp}{lop}
\floatname{algo}{Algorithm}

\def\Bmp#1{ \begin{minipage}{#1} }
\def\Emp{ \end{minipage} }
\def\Bmpc#1{ \begin{minipage}[c]{#1} }
\def\Bmpt#1{ \begin{minipage}[t]{#1} }
\def\Bmpb#1{ \begin{minipage}[b]{#1} }

\def\RR{\mathbb{R}}

\def\J{{\mathcal{J}}}

\def\O{{\mathcal{O}}}

\def\g{{\mathbf{g}}}
\def\bG{{\mathbf{G}}}
\def\b{{\mathbf{b}}}
\def\x{{\mathbf{x}}}

\def\n{{\mathbf{n}}}
\def\t{{\mathbf{t}}}

\def\v{{\mathbf{v}}}

\def\0{{\mathbf{0}}}
\def\bA{{\mathbf{A}}}
\def\bB{{\mathbf{B}}}
\def\bG{{\mathbf{G}}}

\def\tg{\tilde{\Gamma}_{LG}}
\def\tOmega{\tilde{\Omega}}

\def\tF{{\tilde{F}}}

\def\bnabla{\boldsymbol{\nabla}}
\def\bsigma{\boldsymbol{\sigma}}

\def\Dpartial#1#2{ {\partial #1 \over \partial #2} }

\newcommand{\argmin}{\operatorname{argmin}}

\newcommand{\supp}{\operatorname{supp}}

\newcommand{\vp}{\mathbf{v}}

\newcommand{\sig}{\boldsymbol{\sigma}}
\newcommand{\avp}{\langle \mathbf{v} \rangle }

\newcommand{\asigL}{ \langle \boldsymbol{\sigma}_L^{\mu} \rangle}
\newcommand{\asigG}{ \langle \boldsymbol{\sigma}_G^{\mu} \rangle}

\newcommand{\asigLmu}{ \langle \boldsymbol{\sigma}_L^{\mu} \rangle}
\newcommand{\asigGmu}{ \langle \boldsymbol{\sigma}_G^{\mu} \rangle}
\newcommand{\asigmu}{ \langle \boldsymbol{\sigma}^{\mu} \rangle}
\newcommand{\app}{ \langle p \rangle}

\newtheorem{assume}[]{Assumption}

\begin{document}

\title{\vspace*{-1.0cm}
\bf{  Computation of Effective  Free Surfaces in Two Phase Flows}}

\author{R.\ Yapalparvi}
\email{ramesh.yapalparvi@gmail.com}
\author{ B.\ Protas }
 \email{bprotas@mcmaster.ca}
\affiliation{Department of Mathematics \& Statistics, McMaster University \\
 Hamilton, Ontario, Canada.}
\date{\today}

\begin{abstract}
In this investigation we revisit the concept of ``effective free
surfaces'' arising in the solution of the time--averaged fluid
dynamics equations in the presence of free boundaries. This work is
motivated by applications of the optimization and optimal control
theory to problems involving free surfaces, where the time--dependent
formulations lead to many technical difficulties which are however
alleviated when steady governing equations are used instead. By
introducing a number of precisely stated assumptions, we develop and
validate an approach in which the interface between the different
phases, understood in the time--averaged sense, is sharp. In the
proposed formulation the terms representing the fluctuations of the
free boundaries and of the hydrodynamic quantities appear as boundary
conditions on the effective surface and require suitable closure
models. As a simple model problem we consider impingement of
free--falling droplets onto a fluid in a pool with a free surface, and
a simple algebraic closure model is proposed for this system. The
resulting averaged equations are of the free--boundary type and an
efficient computational approach based on shape optimization
formulation is developed for their solution. The computed effective
surfaces exhibit consistent dependence on the problem parameters and
compare favorably with the results obtained when the data from the
actual time--dependent problem is used in lieu of the closure model.
\end{abstract}
\pacs{}
\keywords{Averaged Equations, Free--surface flows, Closure models, 
Volume of Fluid, Shape optimization}
\maketitle

\section{Introduction}
\label{sec:effective_surf_intro}

The goal of this work is to investigate the concept of ``effective
free surfaces'' which are defined here as stationary interfaces
corresponding to the time--averaged balance of mass, momentum and, if
applicable, energy in a time--dependent flow with free surfaces.  In
other words, given an unsteady two--phase flow with fluctuating
boundaries, the effective free surface represents the boundary between
the two phases in the corresponding mean flow which satisfies the
time--averaged form of the original system of governing equations
subject to a number of modeling assumptions.  The motivation for this
work comes from the field of flow control \cite{g03} where many
emerging applications involve control and optimization of
free--surface phenomena. The particular applications underlying this
research concern optimization of complex thermo--fluid phenomena
occurring in liquid metals during welding, see Volkov \emph{et
  al.}\cite{vplg09a}.  While the mathematical foundations for the
optimal control of time--dependent free--boundary problems are
relatively well understood \cite{mz06}, such approaches tend to result
in computational problems of significant complexity even for simple
models \cite{pl08}. The main difficulty arising when methods of the
optimal control, or more broadly, the calculus of variations are
applied to such problems is that some of the optimality conditions
have the form of partial differential equations (PDEs) defined on
interfaces which evolve with time. Needless to say, such problems tend
to be quite hard to solve for non--trivial applications. On the other
hand, this framework becomes much more tractable when time--{\em
  independent} free--boundary problems are considered instead
\cite{hm03}. Moreover, on a more practical level, fluid flows with
free surfaces may generate ``subgrid--scale'' features which are
particularly difficult to compute, and it is therefore desirable to
account for their effect in the average balance of mass and momentum
in a systematic manner. In this paper we propose and test a simple
mathematical model, in the form of a system of coupled PDEs of the
free--boundary type, representing the time--averaged conservation of
mass and momentum in a given time--dependent problem with free
surfaces. While such averaging approaches are well--established in the
study of turbulent flows in domains with fixed boundaries, giving rise
to the well--known Reynolds--Averaged Navier--Stokes (RANS) equations,
see, e.g., Pope\cite{pope}, the additional complication in the present
problem is that one also needs to take into account the effect of the
fluctuations of the location of the free surfaces on the average mass
and momentum balance. Our approach to this problem relies on a number
of simplifying assumptions which are all clearly identified. In the
spirit of the ``closure problem'' arising in turbulence modeling, see
Ref.~\onlinecite{pope}, in order to close the resulting system of
equations one needs to express average products of fluctuating
quantities in terms of average quantities.  However, in contrast to
the classical closure problem where the Reynolds stresses are modeled
with terms defined in the bulk of the fluid, in the present problem,
subject to certain assumptions, such closure terms will appear in the
boundary conditions defined on the effective free boundary. We will
also discuss some very simple strategies for constructing such
closures.  The question of ensemble--averaged, or time--averaged,
description of flows with interfaces has received some attention in
the literature and we mention here the work of Dopazo \cite{d77}, Hong
\& Walker \cite{hw00} and Brocchini \& Peregrine \cite{bp01a,bp01b}
which also contains references to a number of earlier attempts. These
problems were recently revisited in the context of the derivation and
validation of suitable models for multiphase Reynolds--averaged
Navier--Stokes (RANS) equations\cite{b02} and Large Eddy Simulation
(LES) \cite{lltlvlcs07,tlls08,vlltls08}. We also mention the recent
investigation by Wac{\l}awczyk \& Oberlack\cite{wo11} where a number of
closure strategies were proposed for this type of flows. Finally, we
add that the related question of homogenization of free--boundary
problems is an emerging topic in the mathematical analysis of PDEs,
see, e.g., Schweizer \cite{s00}. A detailed description of various
computational methods applied to multiphase flows can be found in the
monograph by Prosperetti \& Tryggvason \cite{prosperetti}. As compared
to these earlier studies, novel aspects of the present investigation
are that, first, we want to compute steady--state solutions, which is
motivated by the optimization applications mentioned above, and
secondly, we want our averaged flows to feature {\em sharp} effective
surfaces, so that the free--boundary property of the original problem
is preserved in its averaged version. In contrast, we note that the
formulations developed in Refs.~\onlinecite{bp01b} and
\onlinecite{wo11} lead to interfaces, referred to as ``surface
layers'', characterized by finite thickness. We also wish to highlight
that although Brocchini \& Peregrine \cite{bp01b} derived averaged
equations taking into account the fluctuations of the free boundaries
and also proposed a simple closure model, to the best of our
knowledge, there have been no attempts to actually compute such
solutions for nontrivial problems which is one of the contributions of
the present work.

\begin{figure}
\centering
\includegraphics[width=0.5 \textwidth]{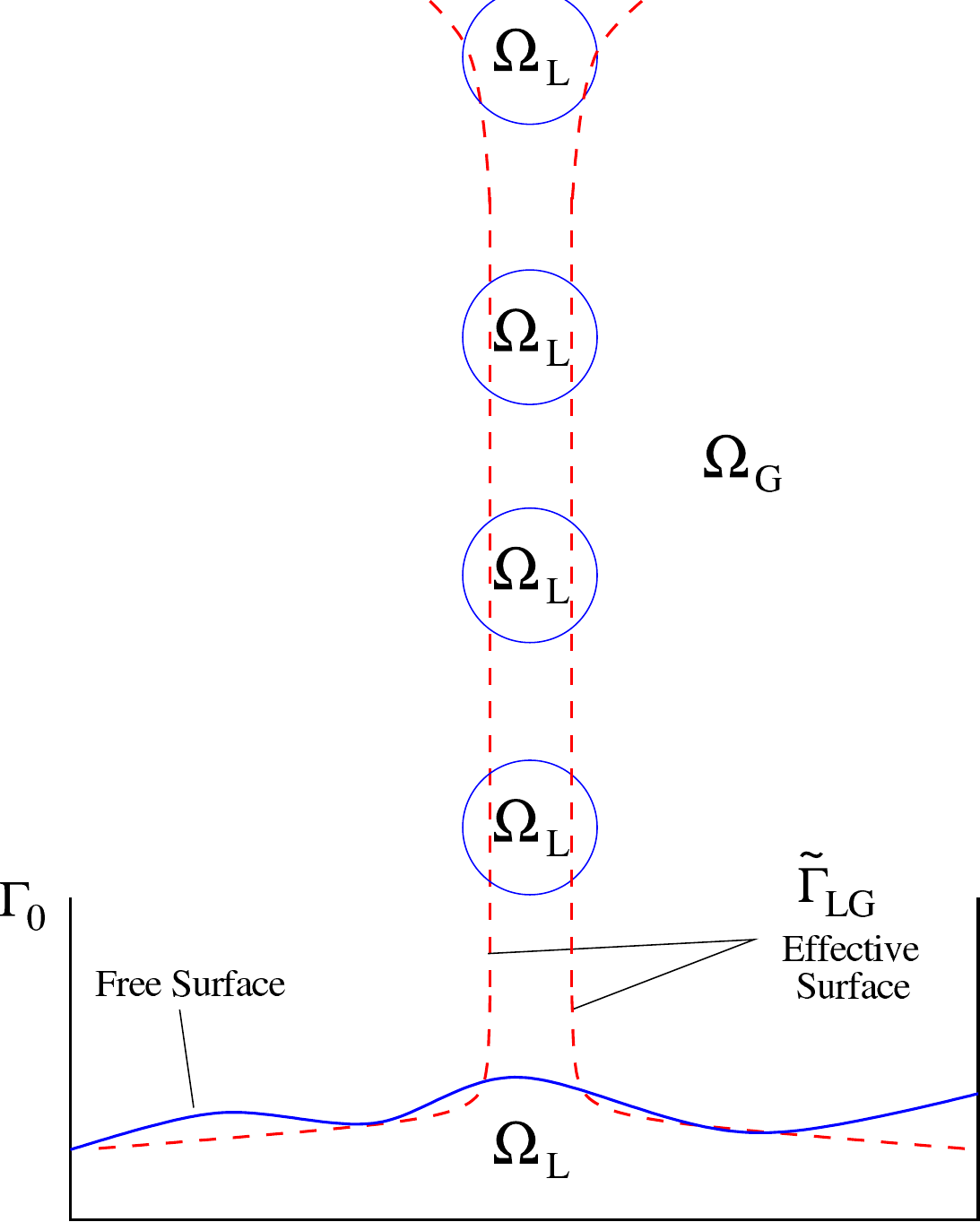} 
\caption{Schematic of our model problem representing droplets
  impinging on a free surface, a phenomenon typically encountered in
  various welding processes such as Gas--Metal Arc Welding (GMAW). The
  solid lines represent the actual time--dependent interface between
  the liquid and gas phases, whereas the dashed line is the steady
  effective surface $\tg$ we seek to determine.}
\label{fig:model}
\end{figure}

The resulting system of PDEs represents the averaged balance of mass
and momentum which has the form of a steady--state free--boundary
problem. Since such problems tend to be difficult to solve
numerically, we also propose a solution approach based on shape
optimization which is well adapted to the numerical solution of this
class of problems. In order to test our approach we choose a very
simple model problem which, while allowing us to focus on certain
methodological aspects, still captures some essential features of the
motivating application, namely, the transfer of mass and momentum to
the weld pool via droplets, see Figure \ref{fig:model}. This model
describes the two--dimensional (2D), time--periodic impingement of
droplets on the free surface of the fluid in a container. In view of
the comments made above, we see that formulation of an optimal control
problem for the original time--dependent system would require us to
satisfy certain optimality conditions on the boundary of each
individual moving droplet in addition to conditions on the free
surface of the liquid in the pool. On the other hand, the concept of
the effective free surface allows us to replace this optimization
problem with a simpler one, which is also computationally more
tractable, where the optimality conditions have to be imposed on the
stationary effective surfaces. Thus, as one application, the proposed
approach will allow us to extend the optimization formulation
developed in Volkov \emph{et al.}\cite{vplg09a} to include the effect
of the mass transfer into the weld pool via droplet impingement.

We remark that droplet impingement onto a thin liquid film is a
phenomenon with manifold manifestations in technology, including
chocolate processing, spray painting, corrosion of turbine blades,
fuel injection in internal combustion engines, and aircraft icing. It
also occurs in many natural phenomena such as the erosion of soil and
the dispersal of spores and micro-organisms.  A considerable amount of
literature is available as concerns the numerical modeling of droplet
impingement onto a solid surface.  Harlow \& Shannon \cite{harlow}
were the first to simulate this phenomenon and several other authors
have applied the Volume--of--Fluid (VoF) based approaches such as
RIPPLE \cite{ripple} and SOLA--VOF \cite{hn81} to understand
droplet impingement phenomena.  Trujillo \emph{et al.}
\cite{trujillo} also performed a numerical investigation and
experimental characterization of the heat transfer from a periodic
impingement of droplets.

The structure of this paper is as follows: in the next Section we
present the formulation of the problem in general terms, in the
following Section we introduce our model problem and in Section
\ref{sec:efs_closure} we discuss a very simple closure strategy which
may be suitable for this problem, in Section \ref{sec:efs_steady} we
introduce a shape--optimization approach to the numerical solution of
the resulting averaged equations, whereas in Section
\ref{sec:efs_results} we present some computational results together
with a discussion; final conclusions are deferred to Section
\ref{sec:final} and some technical results concerning solution of the
shape optimization problem in Section \ref{sec:efs_steady} are
collected in Appendix
\ref{sec:efs_grad}.

\section{Problem Formulation}
\label{sec:problem}

\begin{figure}
\centering
\includegraphics[width=0.5 \textwidth]{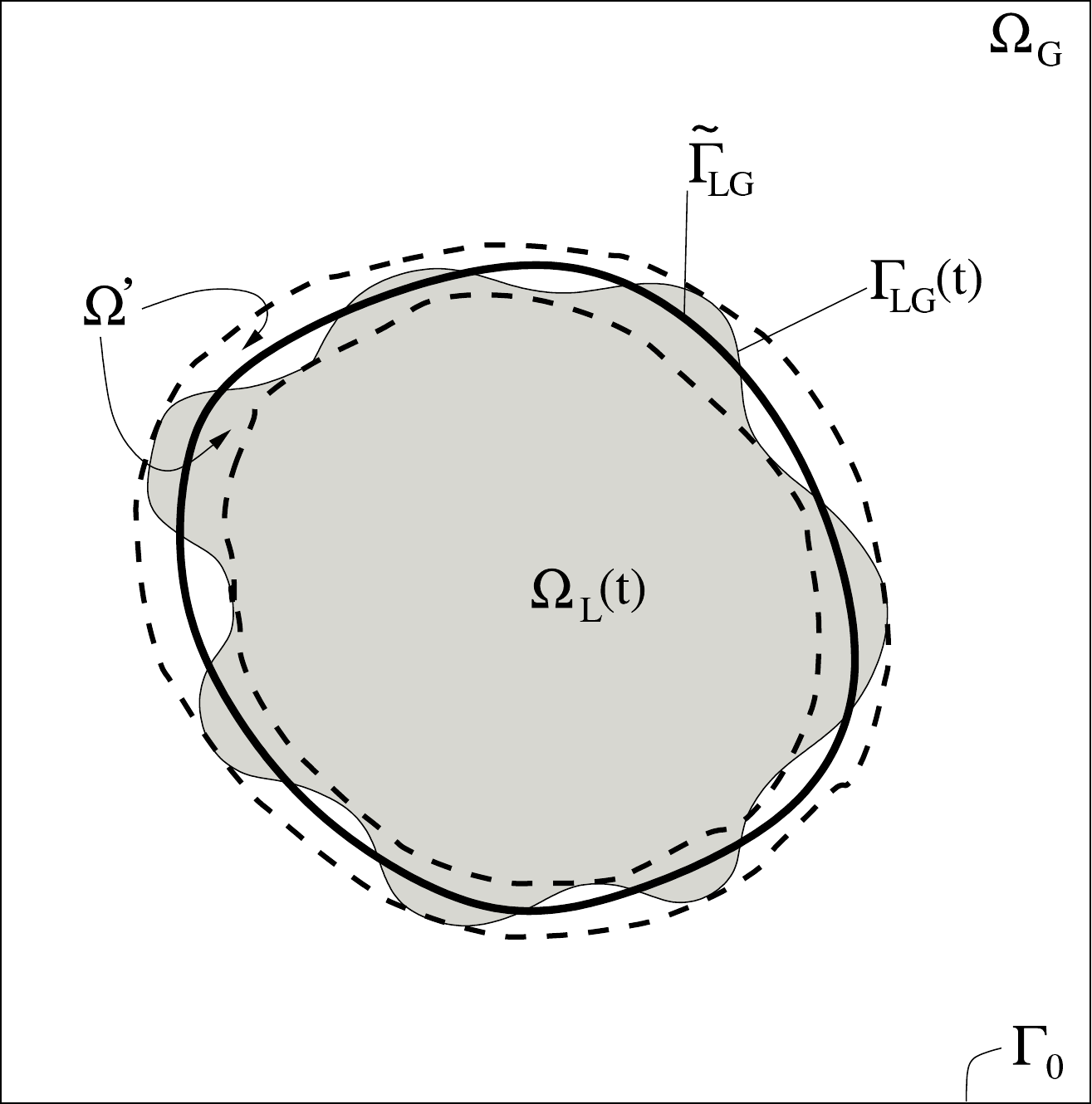} 
\caption{Schematic of the domains and domain boundaries used in the
  definition of the model problem in Section \ref{sec:efs_eqns}.  The
  domain $\Omega_L(t)$ occupied by the liquid phase is marked in gray,
  whereas the thin and thick solid lines represent, respectively, its
  boundary $\Gamma_{LG}(t)$ and the corresponding effective surface
  $\tg$.  The subregion $\Omega'$ (see Section \ref{sec:reduce}) is
  delimited by the thick dashed lines.}
\label{fig:domains}
\end{figure}

In order to simplify the presentation of our approach, we will
consider a two--dimensional problem formulated in a general domain
$\Omega \subset \RR^2$, shown schematically in Figure
\ref{fig:domains}, where $\Omega_L$ and $\Omega_G$ represent the
subdomains occupied, respectively, by the immiscible liquid and gas
phases, whereas $\Gamma_{LG}$ represents the liquid--gas interface
(e.g, droplet boundary or the free surface of the weld
pool). 

\subsection{Assumed Governing Equations}
\label{sec:efs_eqns}

For a general description of the equations and boundary conditions
governing multiphase flows we refer the reader to monograph by
Prosperetti \& Tryggvason\cite{prosperetti}.  We assume that our model
problem involves the following dependent variables
\begin{enumerate}
\item velocity $\textbf{v}= [u,v]^T \, : \, \Omega \times (0,T]
\quad \to  \mathbb{R}^2$,
\item pressure $p \, : \, \Omega  \times (0,T]
\quad \to  \mathbb{R}$ and 
\item position of the free surface 
$\forall_{t \in (0,T]}, \ \Gamma_{LG}(t) \triangleq  \overline{\Omega}_L(t)  \cap  \overline{\Omega}_G(t)$,
\end{enumerate}
where $T$ denotes the length of the time window of interest and
``$\triangleq$'' means ``equal to by definition''.  It is also assumed
that there is no mass transfer across the interface $\Gamma_{LG}$. We
have the following equations governing the fluid flow in the two
phases, for the moment in the time--dependent form
\begin{subequations}\label{ns_liquid}
\begin{align}
\rho_L \Dpartial{\v}{t} + \rho_L (\mathbf{v}\cdot \bnabla) \mathbf{v}- \bnabla
\cdot \sig -
\rho_L \g &=0  \quad \text{in} \  \Omega_L, \label{ns_liquida}\\
\bnabla \cdot \mathbf{v} &=0  \quad \text{in} \  \Omega_L, \label{ns_liquidb} 
\end{align}
\end{subequations}
\begin{subequations}\label{ns_gas}
\begin{align}
 \rho_G \Dpartial{\v}{t} + \rho_G (\mathbf{v}\cdot \bnabla) \mathbf{v} - \bnabla
\cdot \sig -
\rho_G \g &=0  \quad \text{in} \  \Omega_G, \label{ns_gasa}\\
\bnabla \cdot \mathbf{v} &=0  \quad \text{in} \  \Omega_G, \label{ns_gasb}
\end{align}
\end{subequations}
where $\rho_L$ and $\rho_G$ are the densities in the liquid and gas
phase and $\sig \triangleq -p \mathbf{I}+ \sig^{\mu}$ is the stress
tensor in which $\sig^{\mu} \triangleq \mu\big[\bnabla
\mathbf{v}+(\bnabla \mathbf{v})^T\big]$, $\mathbf{I}$ denotes the
identity matrix and the viscosity coefficient $\mu = \mu_L$ or $\mu =
\mu_G$ in the liquid and gas phase, respectively. The symbol $\g$
denotes the gravitational acceleration.  Equations \eqref{ns_liquidb}
and \eqref{ns_gasb} represent conservation of mass, whereas equations
\eqref{ns_liquida} and \eqref{ns_gasa} represent conservation of
momentum in both the liquid and gas phase. Systems \eqref{ns_liquid}
and \eqref{ns_gas} are subject to the following boundary conditions on
the liquid--gas interface $\Gamma_{LG}$
\begin{subequations}\label{bc_interface}
\begin{alignat}{2}
\mathbf{v}\big|_L & =\mathbf{v}\big|_G  \quad &&\textrm{on} \quad \Gamma_{LG}, \label{bc_interface_a} \\
\Big[\sig \Big]_L^G. \mathbf{n} &= \gamma \kappa\, \mathbf{n} \quad
&& \text{on} \quad \Gamma_{LG}, \label{bc_interface_b}
\end{alignat}
\end{subequations}
where $\n$ and $\t$ are the unit normal and tangential vectors on the
interface $\Gamma_{LG}$, $\kappa$ is the interface curvature, $\gamma$
the surface tension (a material property assumed constant), whereas
the subscripts $L$ and $G$ (with or without the vertical bar) denote
quantities defined in the corresponding phases (Figure
\ref{fig:domains}). We note that the vector--valued condition
\eqref{bc_interface_b} implies the balance of both the normal and
tangential stresses. For simplicity, on the far boundary $\Gamma_0$,
cf.~Figure \ref{fig:domains}, we adopt the no--slip boundary condition
\begin{equation}
\v\big|_G = \0 \qquad \textrm{on} \ \Gamma_0.
\label{eq:vg0}
\end{equation}

As regards the mathematical description of free--boundary problems,
there are two main paradigms, namely, (i) ``interface tracking''
approaches, see Neittaanmaki \emph{et al.}  \cite{nst06} and (ii)
``interface capturing'' approaches, see Sethian\cite{s96}. While
description \eqref{ns_liquid}--\eqref{bc_interface}, featuring the
location of the interface $\Gamma_{LG}$ as the dependent variable,
belongs to the first category, for the purpose of developing our
formulation an interface capturing approach will be more suitable and
we employ a technique known as ``Volume of Fluid'' (VoF). However, our
computations of the effective surfaces will be ultimately carried out
with an ``interface tracking'' approach, see Section
\ref{sec:efs_steady}.  A detailed description of the VoF methodology
can be found in the paper by Hirt \& Nichols\cite{hn81}, see also
monograph by Prosperetti \& Tryggvason \cite{prosperetti}.  This
method employs the ``volume fraction'' as an indicator function to
mark different fluids
\begin{equation}\label{F}
\forall_{t \in [0,T]} \qquad 
F(\x,t) = \left \{ \begin{array}{ll}
1 & \x \in \Omega_L \\
0 & \x \in \Omega_G
\end{array} \right..
\end{equation}
While in the continuous setting the interface $\Gamma_{LG}$ is sharp
and the VoF function $F$ may take the values of 0 and 1 only, in a
numerical approximation there may exist a transition region where $0 <
F < 1$ and the fluid can be treated as a mixture of the two fluids on
each side of the interface. The values of the indicator function are
associated with each fluid and hence are propagated as Lagrangian
invariants. Therefore, the indicator function obeys a transport
equation of the form
\begin{equation}\label{transport_F}
\frac{\partial F}{\partial t} + (\vp \cdot \bnabla) F=0 \qquad \textrm{in} \ \Omega.
\end{equation}
Based on the indicator function, local material properties such as the
density $\rho$ of the fluid can be expressed as
\begin{equation}
\rho(F(\x)) = F(\x) \rho_L+ [1-F(\x)] \rho_G. \label{rho_LG}
\end{equation}

Relationship \eqref{rho_LG} allow us to rewrite formulation
\eqref{ns_liquid}--\eqref{bc_interface} in an equivalent form as one
system of conservation equations defined in the {\em entire} flow
domain $\Omega$ where the fluid properties are, in general,
discontinuous across the interface between the two fluids. In this
single--field representation the two fluids are identified by
indicator function \eqref{F}, whereas the material properties are
expressed as piecewise constant functions and can be written in terms
of their values on either side of interface $\Gamma_{LG}$,
cf.~\eqref{rho_LG}.
\begin{subequations}\label{ns_domain}
\begin{alignat}{2}
\Dpartial{\rho(F)}{t} + \bnabla \cdot\left[\rho(F) \, \vp\right] &=0 && 
\text{in} \  \Omega, \label{ns_domain_a}\\ 
\frac{\partial \rho(F)\, \vp}{\partial t}+\bnabla \cdot\left[\rho(F) \, \vp \vp\right]&=
\bnabla \cdot \sig + \int_{\Gamma_{LG}} \gamma \kappa\, \n \delta(\x-\x')
\, ds(\x')  \quad && \text{in} \  \Omega, \label{ns_domain_b} \\
\frac{\partial F}{\partial t} + (\vp \cdot \bnabla) F &=0 && \textrm{in} \ \Omega, \label{ns_domain_c}
\end{alignat}
\end{subequations}
where $\vp \vp$ denotes the dyadic product, i.e., the tensor defined
as $\left[ \vp \vp \right]_{ij} = \left[ \vp \right]_{i}\left[ \vp
\right]_{j}$, $i,j=1,2$. Conservation equations \eqref{ns_domain_a}
and \eqref{ns_domain_b} can be obtained in a straightforward manner by
considering the integral balance of mass and momentum for the fluid
with variable density $\rho(F)$ in some arbitrary control volume.
Further discussion of the single--field description of two--phase
flows can be found in the monograph by Prosperetti \& Tryggvason
\cite{prosperetti}.

The last term on the right--hand side (RHS) in \eqref{ns_domain_b}
represents the source of momentum due to the surface tension. It is
related to boundary condition \eqref{bc_interface_b} and only acts at
the interface $\Gamma_{LG}$ as indicated by the presence of the Dirac
delta function in the integrand expression of the integral.  The
surface integral in equation \eqref{ns_domain_b} can be difficult to
evaluate directly. In order to overcome this problem, a continuum
surface force (CSF) model was introduced by Brackbill \emph{et al.}
\cite{bkz92} which represents the surface tension effects in terms of
a continuous volumetric force acting within the transition region
which arises when the problem is discretized.  The surface integral in
\eqref{ns_domain_b} is therefore approximated as in Prosperetti and
Tryggvason\cite{prosperetti}
\begin{equation}\label{csf}
\int_{\Gamma_{LG}} \gamma \kappa'\, \n' \delta(\x-\x')
d\Gamma \approx \gamma \kappa \, \bnabla F, 
\end{equation}
whereas the curvature of the interface can be computed in terms of the
VoF function as follows
\begin{equation}\label{kappa}
\kappa= \bnabla \cdot \left( \frac{ \bnabla F}{\mid\bnabla F \mid}\right).
\end{equation}
Using \eqref{csf} in \eqref{ns_domain_b}, we obtain a simpler form of
the one--field system \eqref{ns_domain}, namely
\begin{subequations}\label{ns_csf}
\begin{alignat}{2}
\Dpartial{\rho(F)}{t} + \bnabla \cdot\left[\rho(F) \, \vp\right] &=0 && 
\text{in} \  \Omega, \label{ns_csf_a}\\ 
\frac{\partial \rho(F)\, \vp}{\partial t}+\bnabla \cdot\left[\rho(F)\, \vp \vp\right]&=
\bnabla \cdot \sig + \gamma \kappa\, \bnabla F  \quad 
&& \text{in} \  \Omega, \label{ns_csf_b} \\
\frac{\partial F}{\partial t} + (\vp \cdot \bnabla) F & = 0 && \textrm{in} \ \Omega. \label{ns_csf_c}
\end{alignat}
\end{subequations}

\subsection{Averaging Procedures}
\label{sec:efs_avr}

The goal of this Section is to derive a time--averaged form of
governing system \eqref{ns_liquid}--\eqref{bc_interface}, or
equivalently \eqref{ns_csf}, and state the ``closure problem'', i.e.,
identify the terms in the resulting equations which need to be
modeled. Our objective is to express the averaged equations solely in
terms of averaged velocity, averaged pressure and averaged indicator
function as the dependent variables. A number of different averaging
techniques have been considered in the literature in regard to
multiphase flows \cite{d77,hw00,bp01b,wo11,prosperetti}. Here we will
rely on the conventional time--averaging procedure, see Monin \&
Yaglom \cite{my71} which is based on the ideas originally due to
Reynolds (it should be added that in statistical physics averaging is
typically performed with respect to realizations, however, in view of
the ergodicity assumption adopted here, the ensemble average can be
replaced with a time average used in \eqref{def_average} below). Given
the quantity $\varphi \: : \: [0,T] \times \Omega \rightarrow \RR^d$,
$d=1,2$, we thus define the pointwise time average as
\begin{equation}\label{def_average}
\langle \varphi \rangle(\x) \triangleq 
\frac{1}{\Delta t} \int_{t_0}^{t_0+\Delta t} \varphi(t,\x)\,dt,
\end{equation}
where the time window $\Delta t$ is assumed large compared to the time
scale of the random fluctuations associated with free boundaries.
Since in the present problem we are interested in steady solutions, we
take $\Delta t \rightarrow \infty$, so that the averaged variables do
not depend on time. In the conventional Reynolds decomposition, the
chaotically varying flow variables are replaced by the sums of their
time averages and fluctuations, i.e.,
\begin{equation}\label{variables_average}
\vp = \langle \vp \rangle+\vp', \quad \rho=\langle \rho \rangle+
\rho', \quad p=\langle p \rangle+p'.
\end{equation}
By definition, the time average of a fluctuating quantity is zero,
i.e., $\langle \vp' \rangle \equiv 0$, $\langle \rho' \rangle \equiv
0$ and $\langle p' \rangle \equiv 0$.  We also note that the averaging
operator $\langle \cdot \rangle$ commutes with differentiation with
respect to the space variables \cite{my71,d77}. We shall furthermore
assume that \cite{my71}
\begin{equation}
\left\langle \Dpartial{\varphi'}{t} \right\rangle = 0.
\label{eq:dv'dt}
\end{equation}

Our derivation of the averaged equation follows the general
development presented in Hong \& Walker \cite{hw00}, although we use a
somewhat different notation adapted to the present problem. We begin
with continuity equation \eqref{ns_domain_a} and decompose the
dependent variables as in \eqref{variables_average}. The equation is
then time--averaged and we obtain
\begin{equation}\label{average_continuity}
\bnabla \cdot (  \langle \rho \rangle \langle \vp \rangle) = 
- \bnabla \cdot (\langle \rho' \vp' \rangle).
\end{equation}
We need to re--express the right hand side of equation
\eqref{average_continuity} to eliminate $\rho'$. From \eqref{rho_LG}
and applying the Reynolds decomposition to the indicator function
\eqref{F}
\begin{equation}
F(t,\x) = \langle F \rangle(\x) + F'(t,\x), \quad \textrm{where} \ 
\langle F'\rangle(\x) \equiv 0, 
\label{eq:Fr}
\end{equation}
we obtain
\begin{equation}\label{rho}
\begin{aligned}
\rho(t,\x) & = [\langle F \rangle(\x) + F'(t,\x)] \rho_L + 
[1-\langle F \rangle(\x) - F'(t,\x) ] \rho_G \\
& = \langle \rho \rangle+ F' (t,\x)(\rho_L-\rho_G)
\end{aligned}
\end{equation}
which allows us to identify $\rho'(t,\x) = F'(t,\x) \,
(\rho_L-\rho_G)$.  Using \eqref{rho} we can now deduce
\begin{equation}\label{average_rho'v'}
\langle \rho' \vp' \rangle= (\rho_L-\rho_G) \langle F' \vp' \rangle,
\end{equation}
so that \eqref{average_continuity} becomes
\begin{equation}\label{continuity_f'v'}
\bnabla \cdot \Big\{ \left[\rho_L \langle F\rangle  (\x)+ 
\rho_G (1 - \langle F\rangle (\x)) \right] \langle \vp \rangle) \Big\} = 
 - (\rho_L-\rho_G)\bnabla \cdot \langle F' \vp' \rangle.
\end{equation}
which is the Reynolds--averaged form of the continuity equation, where the
right--hand side (RHS) terms represent the average effect of the
fluctuations of the free boundary.

We now turn our attention towards momentum equation \eqref{ns_csf}.
In order to simplify the formulation of the present problem we make
the following
\begin{assume}
  The fluctuations of viscosity $\mu(t,\x) = \mu_L F(t,\x) + \mu_G
  [1-F(t,\x)]$ and the interface curvature $\kappa$,
  cf.~\eqref{kappa}, are neglected. These quantities will be therefore
  treated as constant and will not be subject to averaging.
\label{assume1a}
\end{assume}
\noindent The reason for this simplification is that proper handling
of viscosity and curvature fluctuations leads to significant
complications in the resulting average equations, and this issue is
deferred to future research. In a phenomenon characterized by an
interplay of capillary, viscous and inertial effects, Assumption
\ref{assume1a} implies that the applicability of the model is
restricted to flow regimes dominated by the inertial effects.  Indeed,
we expect that the density fluctuations are going to have a dominating
influence on the effective surfaces in the class of applications
motivating this study. The development of the Reynolds--averaged form
of the momentum equations proceeds most easily when the nonlinear
advection terms are written in the conservative form. Again, at every
point $\x$ we replace the dependent variables with relations
\eqref{variables_average} and then average the equations over time.
The complete Reynolds--averaged momentum equation can be written as
\begin{equation}
  \bnabla \cdot\Big(
\langle \rho  \rangle  \langle \vp \rangle \langle \vp\rangle \Big) = 
- \bnabla \langle p  \rangle + \bnabla \cdot\Big(
\langle \bsigma^{\mu}   \rangle  - \langle \rho  \rangle \langle \vp' \vp' \rangle 
- \langle \vp \rangle  \langle \rho' \vp' \rangle 
- \langle \rho'  \vp' \vp' \rangle \Big)+ \gamma \kappa\, \bnabla \langle F\rangle, 
\label{momentum}
\end{equation}
where $\langle \bsigma ^{\mu} \rangle$ is the usual viscous stress
tensor defined in terms of the averaged velocity field and, in
addition to the Reynolds stresses, on the RHS in \eqref{momentum} we
also note the presence of new terms representing fluctuations of the
free boundaries.

The main idea behind the proposed approach is that the resulting
averaged solutions should preserve some essential features of the
original time--dependent free--boundary problem
\eqref{ns_liquid}--\eqref{bc_interface}, namely, a sharp separation
between the two phases, cf.~Figure \ref{fig:model}, along an interface
which we defined as the effective free surface. While the
time--dependent indicator function $F$ may only assume the values of 0
and 1, cf.~\eqref{F}, its average $\langle F\rangle$ may assume all
intermediate values $0 \le \langle F\rangle(\x) \le 1$ (this is in
fact clearly visible in the plots of the mean indicator function
$\langle F \rangle$ obtained by averaging the solutions of our model
problem, see Figure \ref{figavg}a to be discussed further below in
Section \ref{sec:efs_closure}). Such smoothly varying indicator
functions $\langle F \rangle$ correspond to a continuous transition
between the two phases without a well--defined interface.  Therefore,
in order to be able to define averaged flows with {\em sharp}
effective boundaries we have to introduce the following
\begin{assume}
The average indicator function $\langle F \rangle$ is replaced in
Reynolds--averaged equations \eqref{continuity_f'v'}--\eqref{momentum}
with the piecewise--constant function $\tF \; : \; \Omega \rightarrow
\RR$ defined as follows
\begin{equation}\label{tF}
\tF(\x) = \left \{ \begin{array}{ll}
1, & \x \in \tOmega_L \\
0, & \x \in \tOmega_G
\end{array} \right.,
\end{equation}
where $\tOmega_L$ and $\tOmega_G$ are the corresponding
time--invariant subdomains occupied by the liquid and gas phases.
\label{assume1}
\end{assume}
\noindent With this assumption the Reynolds--averaged equations take the form
\begin{subequations}\label{nsR}
\begin{alignat}{2}
& \bnabla \cdot \Big\{ \left[\rho_L \tF(\x) +  \rho_G (1 - \tF(\x)) \right] \langle \vp \rangle) \Big\} =   - \bnabla \cdot \bA && 
\text{in} \  \Omega\label{nsRa}\\ 
& \bnabla \cdot\Big(
\langle \rho  \rangle  \langle \vp \rangle \langle \vp\rangle \Big) = 
- \bnabla \langle p  \rangle + \bnabla \cdot\Big(
\langle \bsigma^{\mu} \rangle - \langle \rho  \rangle  \langle \vp'\vp' \rangle + \bB \Big)+ \gamma \kappa\, \bnabla \tF, \quad && \text{in} \  \Omega \label{nsRb}
\end{alignat}
\end{subequations}
where the vector $\bA$ and tensor $\bB$ are defined as
\begin{subequations}
\label{eq:AB}
\begin{align}
\bA & \triangleq (\rho_L-\rho_G)
\begin{bmatrix}
  \langle F'u' \rangle \\
\langle F'v' \rangle   \\
\end{bmatrix},
\label{eq:A} \\
\mathbf{B} & \triangleq (\rho_L-\rho_G)
\begin{bmatrix}
 2 \langle u \rangle \langle F'v' \rangle+  \langle F'u' u'\rangle &   \langle u\rangle \langle F'v' \rangle+ \langle v\rangle \langle F'u' \rangle+ \langle F'u' v'\rangle\\
\langle u\rangle \langle F'v' \rangle+ \langle v\rangle \langle F'u' \rangle+ \langle F'u' v'\rangle &  2 \langle v \rangle \langle F'u' \rangle+  \langle F'v' v'\rangle  \\
\end{bmatrix}.
\label{eq:B}
\end{align}
\end{subequations}
As one can see, equations \eqref{nsR} are not ``closed'', because they
contain averaged products of fluctuation terms for which no additional
equations are available.  Therefore, we will seek to model these terms
with closure expression of the form $\bA = \bA(\tF,\langle \vp
\rangle)$ and $\bB = \bB(\tF,\langle \vp \rangle)$ which are functions
of the averaged dependent variables $\tF$ and $\langle \vp \rangle = [
\langle u \rangle, \langle v \rangle ]^T$.  This is in addition to the
closures required for the ``classical'' Reynolds stress tensor
$\langle \v' \v' \rangle$. While modeling the latter expressions is a
rather well--advanced area \cite{pope}, development of closures for
product terms corresponding to fluctuations of the free boundaries has
been the subject of relatively few investigations, see, e.g.,
Refs.~\onlinecite{bp01b} and \onlinecite{wo11}, which focused on the
case with a diffuse interface. A very simple closure model for these
terms adapted to the present formulation of the problem with a sharp
interface, cf.~Assumption \ref{assume1}, will be presented in Section
\ref{sec:efs_closure}. The question of closure models for the Reynolds
stress terms will not be considered in this work.

\subsection{Reduction of Averaged  Fluctuation Terms to Boundary Conditions}
\label{sec:reduce}

In the derivation of the closure models the quadratic and cubic
products involving the fluctuation fields $F'$, $u'$ and $v'$ will
need to be expressed {\em solely} in terms of the time--averaged
fields $\tF$, $\langle u \rangle$ and $ \langle v\rangle$. As regards
the dependence on $\tF$, this means that expressions for these
closures will depend on the location relative to the effective free
surface and, evidently, the components of the tensors $\bA$ and $\bB$
are nonvanishing only in a close proximity of the free boundary
$\tilde{\Gamma}_{LG}$, cf.~Figure \ref{fig:domains}. From the point of
view of the formulation of a computation--oriented model it is
therefore not ``economical'' to introduce new terms into the averaged
equations which would be nonzero only in a very small fraction of the
domain. We therefore propose the following simplifying approach in
which the averaged fluctuation terms involving tensors $\bA$ and $\bB$
defined in the bulk are approximated with suitable terms defined on
the effective boundary $\tilde{\Gamma}_{LG}$. This can be done by
integrating the terms involving $\bA$ and $\bB$ in
\eqref{nsRa}--\eqref{nsRb} over their support $\Omega' \triangleq
\supp\bA = \supp\bB$ and then using the divergence theorem (in
principle, the supports of these two terms may in general be
different, but for the sake of simplicity we assume here that they
coincide; this simplification does not in any way affect the accuracy
of the proposed approach). We remark that analogous ideas were also
pursued by Brocchini \& Peregrine \cite{bp01b} and by
Brocchini\cite{b02}.  One important difference between these
approaches and the formulation explored here concerns the description
of the effective boundary (explicit in Refs \onlinecite{bp01b} and \onlinecite{b02}
versus intrinsic considered here).  Noting that the fields $\bA$ and
$\bB$ are discontinuous at the effective surface $\tg$ (which is
contained inside the integrations domain $\Omega'$), and vanish on
$\partial\Omega'$, we obtain
\begin{subequations}
\label{eq:I1}
\begin{alignat}{3}
I_1 &= \int_{\Omega'} \bnabla\cdot\bA \, d\Omega & = & 
\int_{\tg} \left[ \n\cdot\bA\right]_L^G \, d\sigma & = & \int_{\tg} a \, d\sigma,
\label{eq:I1a} \\
I_2 &= \int_{\Omega'} \bnabla\cdot\bB \, d\Omega & = &
\int_{\tg} \left[ \n\cdot\bB\right]_L^G \, d\sigma & = & \int_{\tg} \b \, d\sigma
\label{eq:I1b}
\end{alignat}
\end{subequations}
in which the fields $a \; : \; \tg \rightarrow \RR$ and $\b \; : \;
\tg \rightarrow \RR^2$ are defined in terms of the jumps of $\bA$ and $\bB$ as
\begin{equation}
a \triangleq  \left[ \n\cdot\bA\right]_L^G, \qquad 
\b = \begin{bmatrix} b_1 \\ b_2 \end{bmatrix} \triangleq  \left[ \n\cdot\bB\right]_L^G.
\label{eq:ab}
\end{equation}
We thus see that in the {\em mean sense} the fluxes due to the
fluctuating terms $\bnabla\cdot\bA$ and $\bnabla\cdot\bB$ in the
averaged mass and momentum equations \eqref{nsRa} and \eqref{nsRb} can
be realized by the terms $a$ and $\b$, cf.~\eqref{eq:ab}, defined on
the effective boundary $\tg$. This leads to the following 
\begin{assume}
which has two parts
\begin{enumerate}
\item
we replace the source term $\bnabla\cdot\bA$ in averaged mass
conservation equation \eqref{nsRa} with an additional term $(a\, \n)$ in
the corresponding boundary condition \eqref{bc_interface_a},

\item
we replace the source term $\bnabla\cdot\bB$ in averaged momentum
conservation equation \eqref{nsRb} with an additional term $\b$ in
the corresponding boundary condition \eqref{bc_interface_b},
\end{enumerate}
so that the following system of equations is obtained (rewritten here
in the two subdomains together with all boundary conditions)
\begin{subequations}\label{nsR2}
\begin{alignat}{2}
\rho_L \big(\langle \v \rangle\cdot \bnabla\big) \langle \v \rangle- \bnabla \cdot \langle \sig\rangle -
\rho_L \b &=0  \quad && \textrm{in} \  \tOmega_L, \label{nsR2a} \\
\bnabla \cdot \langle \v \rangle &=0  \quad && \textrm{in} \  \tOmega_L, \label{nsR2b}  \\
\rho_G \big(\langle \v \rangle\cdot \bnabla\big) \langle \v \rangle - \bnabla
\cdot \langle \sig \rangle - \rho_G \g &=0  \quad && \textrm{in} \  \tOmega_G, \label{nsR2c} \\
\bnabla \cdot \langle \v \rangle &=0  \quad && \textrm{in} \  \tOmega_G, \label{nsR2d} \\
\left[\avp\right]_L^G & = a\,\n  \quad && \textrm{on} \ \tilde{\Gamma}_{LG}, \label{nsR2e} \\
\n \cdot \left[\langle \sig \rangle \right]_L^G &= \gamma \kappa\, \mathbf{n} + \b \quad && \textrm{on} \ \tilde{\Gamma}_{LG}, \label{nsR2f} \\
\left(\avp\big|_L +\avp\big|_G \right)  \cdot \n&=0 && \textrm{on} \ \tilde{\Gamma}_{LG}, \label{nsR2g}
\end{alignat}
\end{subequations}
where boundary condition \eqref{nsR2g} corresponds to condition
\eqref{eq:vg0} in the situation when the normal velocity at the
effective surface is allowed to have a discontinuity,
cf.~\eqref{nsR2e}.
\label{assume2}
\end{assume}
\noindent As is evident from Figure \ref{fig:domains}, this Assumption
is satisfied when the subregion $\Omega'$ forms narrow bands along the
effective free boundary $\tilde{\Gamma}_{LG}$ which happens when the
fluctuations of the free boundary $\Gamma_{LG}$ occur at a
length--scale significantly smaller than the characteristic dimension
of the entire domain. As regards the averaged conservation equations,
Assumption \ref{assume2} has the effect of reducing, or localizing,
the influence of the averaged terms involving fluctuation of the free
boundary to the effective free boundary $\tilde{\Gamma}_{LG}$.  The as
of now undefined functions $a = a(\tF, \langle \v \rangle)$ and $\b =
\b(\tF, \langle \v \rangle)$ represent the required closure models and
need to be determined separately for every flow problem. We add, that
since these functions depend on the location of the effective free
surface $\tilde{\Gamma}_{LG}$, boundary conditions \eqref{nsR2e} and
\eqref{nsR2f} are in fact geometrically nonlinear.  We also remark
that in Ref.~\onlinecite{bp01b} closure models for certain free
boundary problems were derived based on an analogous concept of
integral balances in the surface layer. Construction of a very simple
closure model for functions $a$ and $\b$ applicable to a model problem
introduced in the next Section will be presented in Section
\ref{sec:efs_closure}.

\section{Model Problem}
\label{sec:model}

While up to this point our discussion has been concerned with a
generic two--phase, free--boundary problem, we will from now on focus
on a specific flow configuration. Thus, to fix attention, we will
consider the flow set--up shown schematically in Figure
\ref{fig:model}. It features droplets entering the domain $\Omega$
periodically through the top boundary and impinging on the free
surface resulting in sloshing. On the lateral boundaries $\Gamma_0$
no--slip boundary condition \eqref{eq:vg0} is applied and we observe
that, respectively, an unsteady or steady contact line will appear
where the time--dependent interface $\Gamma_{LG}$, or the
corresponding effective surface $\tg$, intersects the boundary
$\Gamma_0$. While it is well known that subject to the classical
no--slip and free--surface boundary conditions the contact--line
problem is not well--posed\cite{s07}, development of a both
mathematically and physically consistent description of this problem
still remains an open question.  Addressing this issue is beyond the
scope of the present investigation, and our treatment of the contact
line is a standard one: in the solution of the time--dependent problem
a suitable regularizing effect is achieved by discretization of the
governing equations (described further below), whereas in the solution
of the steady problem with the effective surface regularization is
introduced via formulation in terms of variational shape optimization.
Application of this numerical approach to a closely related problem
with a contact line singularity is analyzed in detail by Volkov and
Protas\cite{vp08a}. In order to maintain a constant average (over one
period of droplet impingement) mass of the fluid $M \triangleq
\int_{\Omega} F(t,\x)\, d\Omega$, the fluid is drained through the
bottom boundary of the domain (i.e., suitable nonzero velocity
boundary condition $\v\cdot\n \neq 0$ is applied there).

\begin{figure}
\begin{center}
\mbox{
\subfigure[  $\hspace*{0.1cm} t=0.0 \hspace*{1.2cm}$]{\hspace*{-1.75cm}
\includegraphics[width=.4\textwidth]{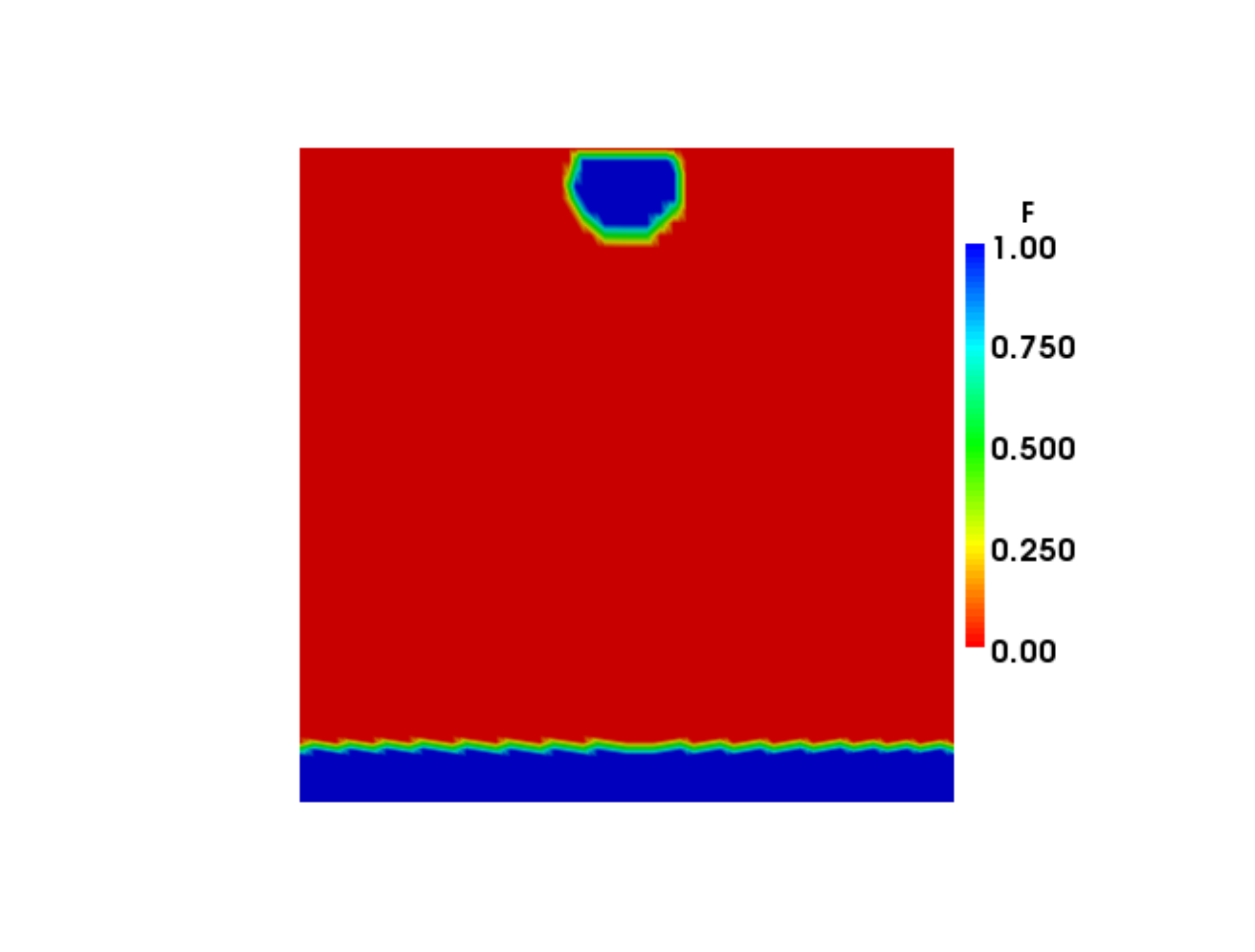}}
\subfigure[  $\hspace*{0.1cm} t=(1/4)T$]{
\includegraphics[width=.4\textwidth]{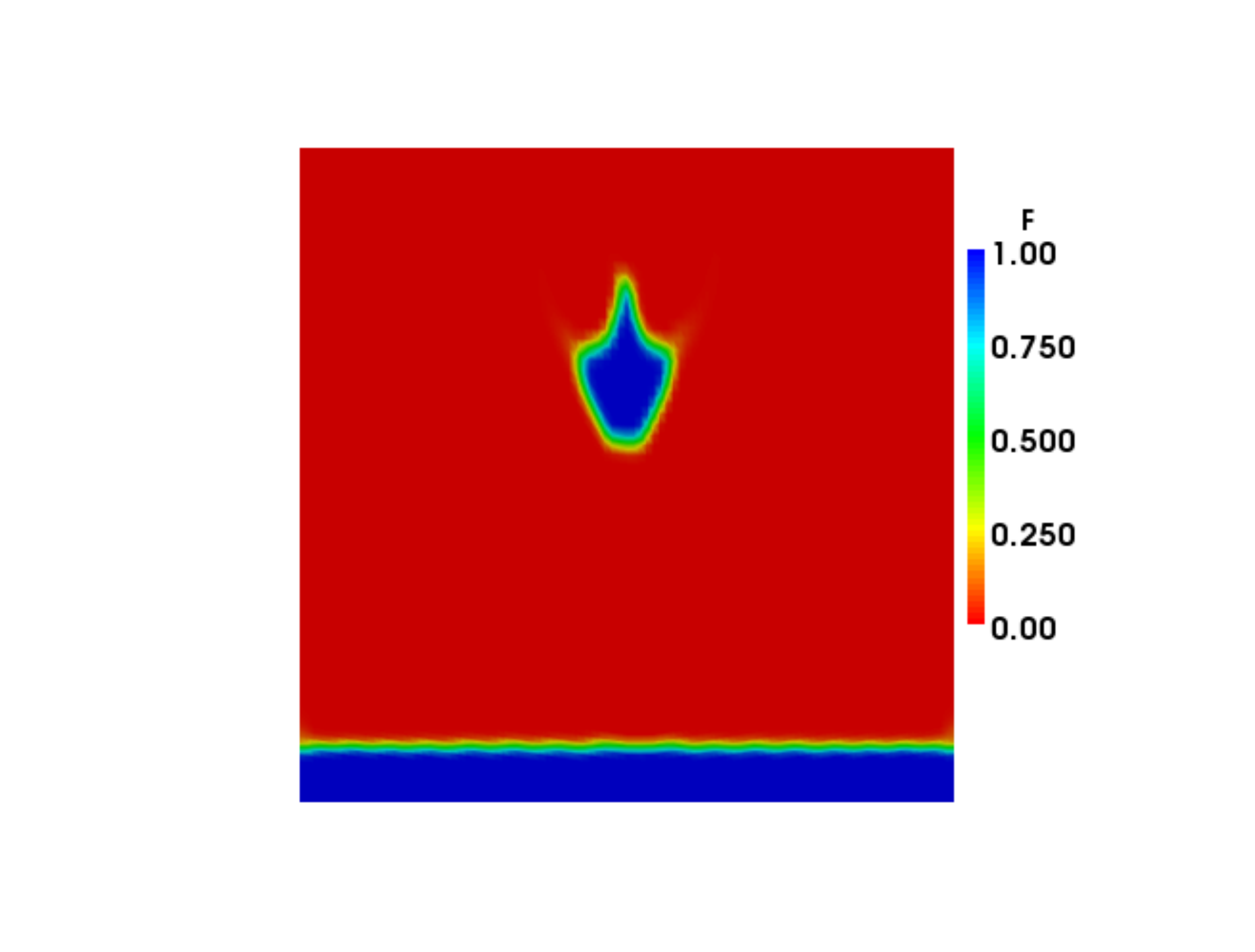}}
\subfigure[$\hspace*{0.1cm} t=(1/2)T$]{
\includegraphics[width=.4\textwidth]{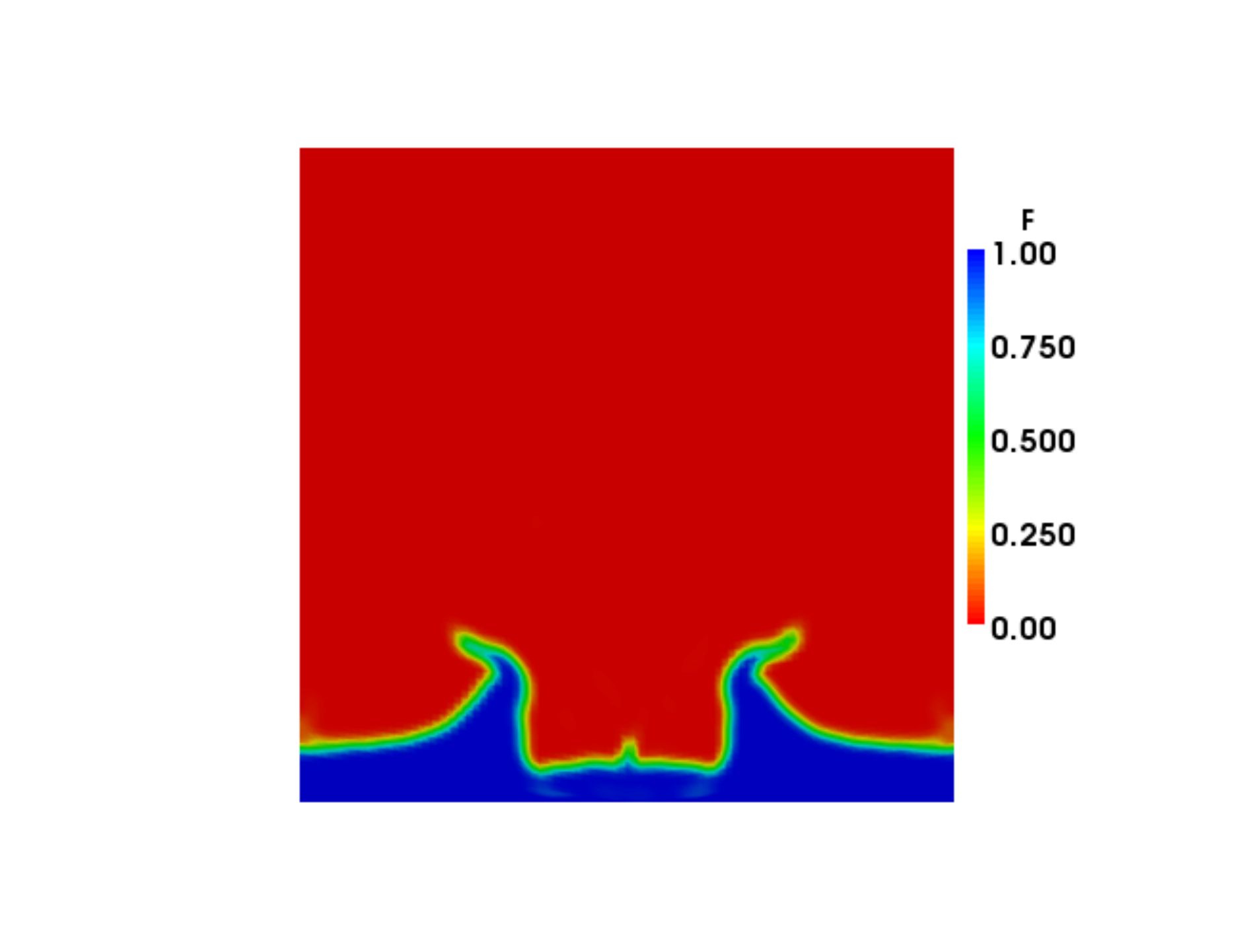}}}
\mbox{
\subfigure[$\hspace*{0.1cm} t=(3/4)T \hspace*{1.2cm} $]{\hspace*{-1.75cm}
\includegraphics[width=.4\textwidth]{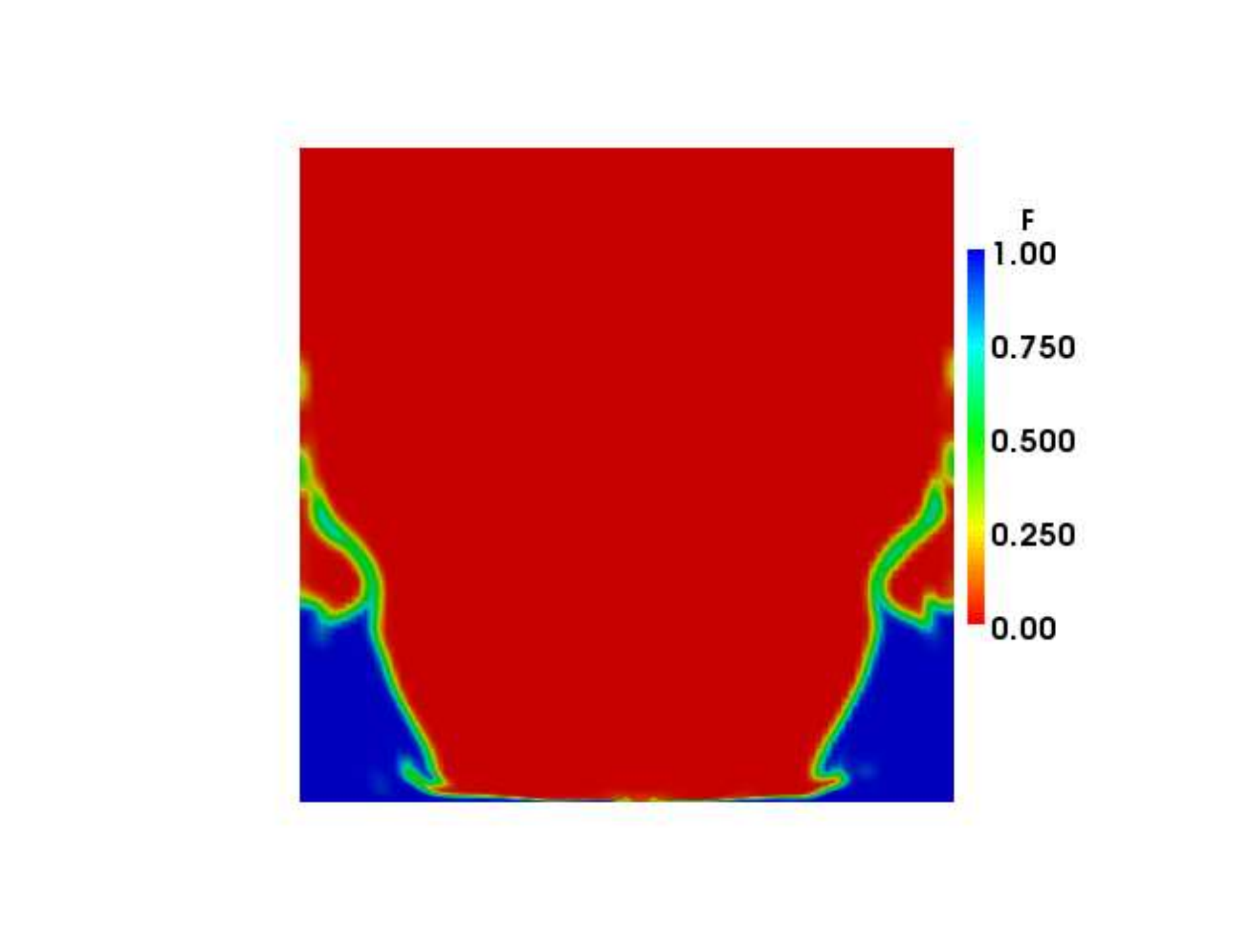}}
\subfigure[$\hspace*{0.1cm} t=T$]{
\includegraphics[width=.4\textwidth]{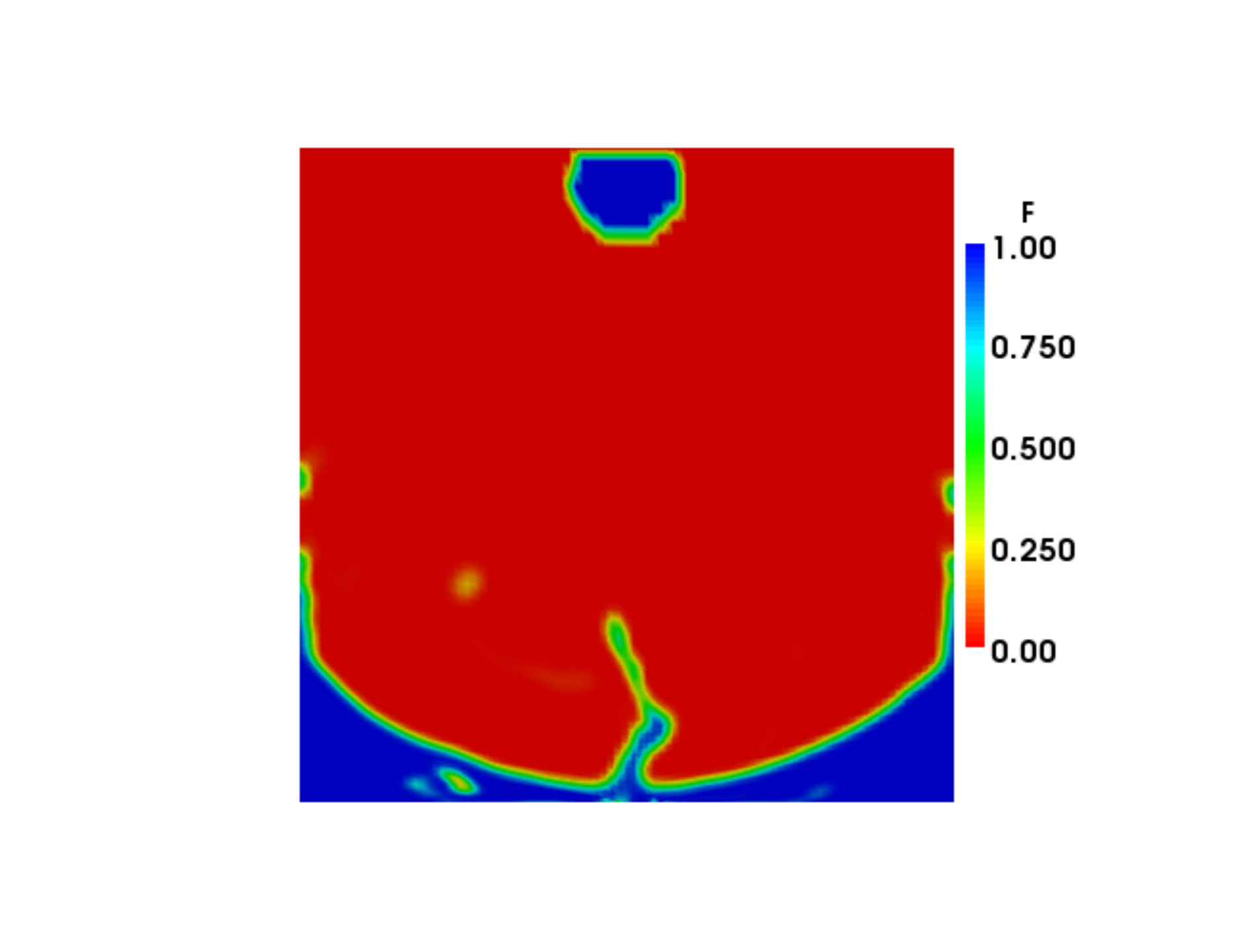}}
\subfigure[$\hspace*{0.1cm} t=(5/4)T$]{
\includegraphics[width=.4\textwidth]{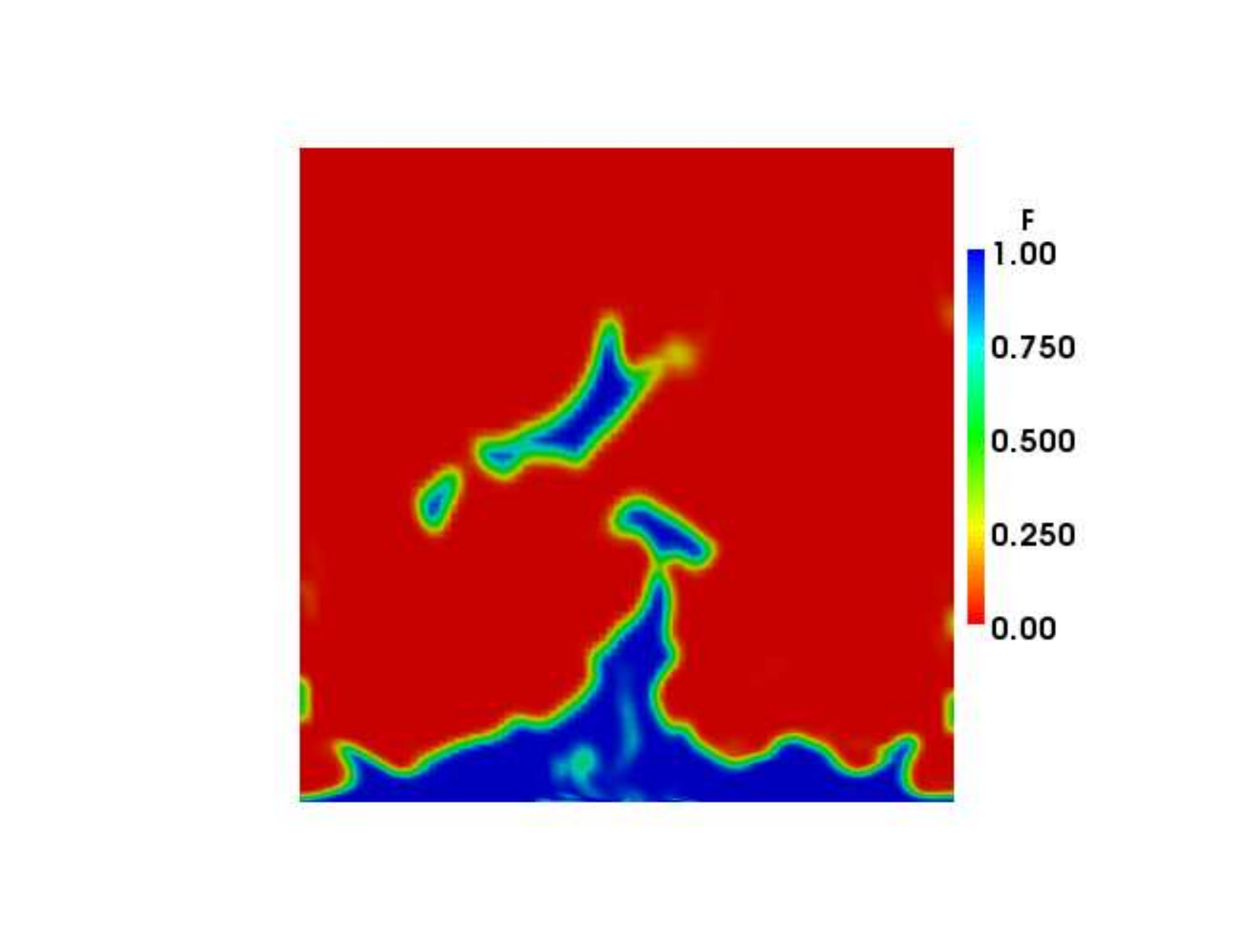}}}
\mbox{
\subfigure[$\hspace*{0.1cm} t=(3/2)T  \hspace*{1.2cm}$]{\hspace*{-1.75cm}
\includegraphics[width=.4\textwidth]{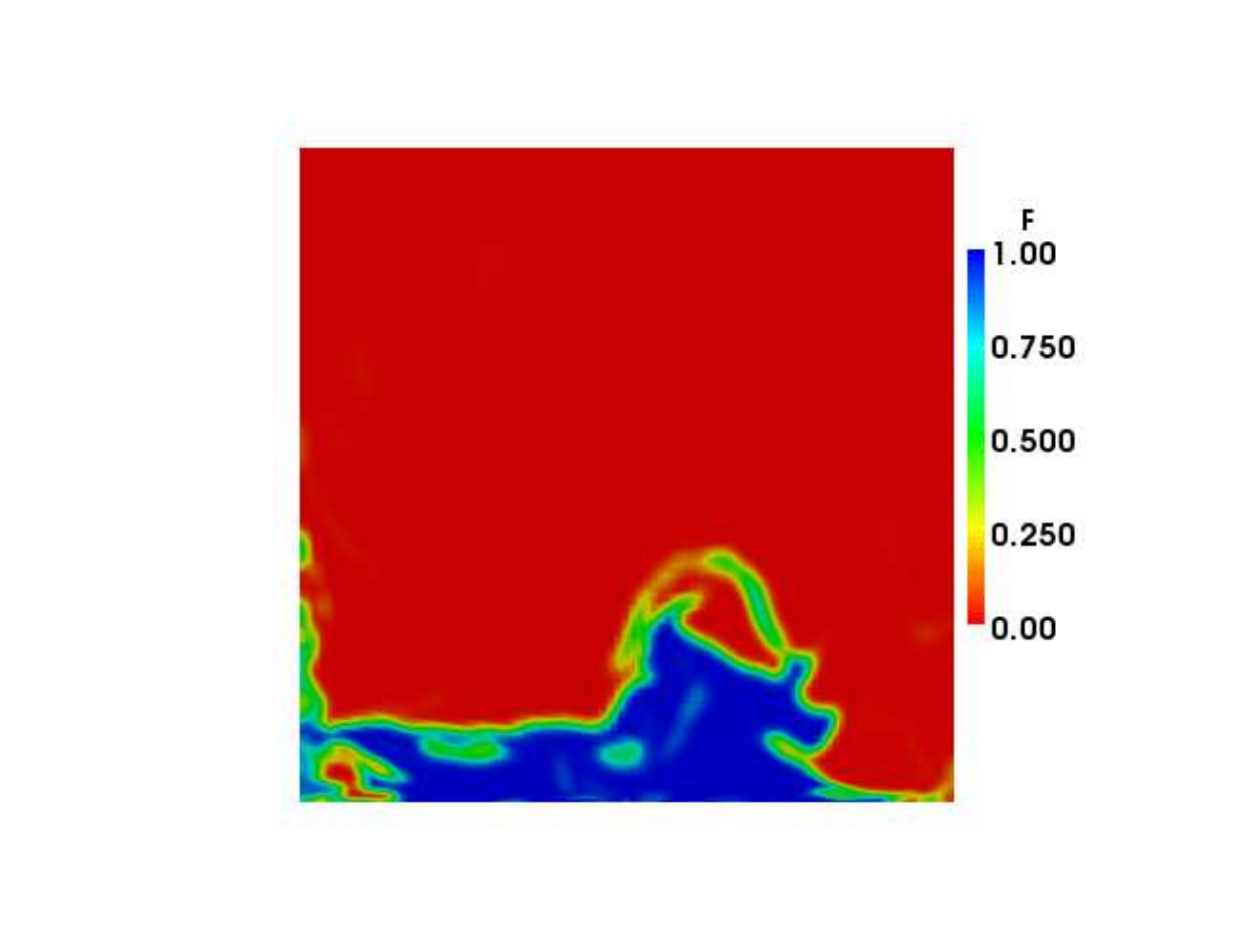}}
\subfigure[$\hspace*{0.1cm} t=(7/4)T$]{
\includegraphics[width=.4\textwidth]{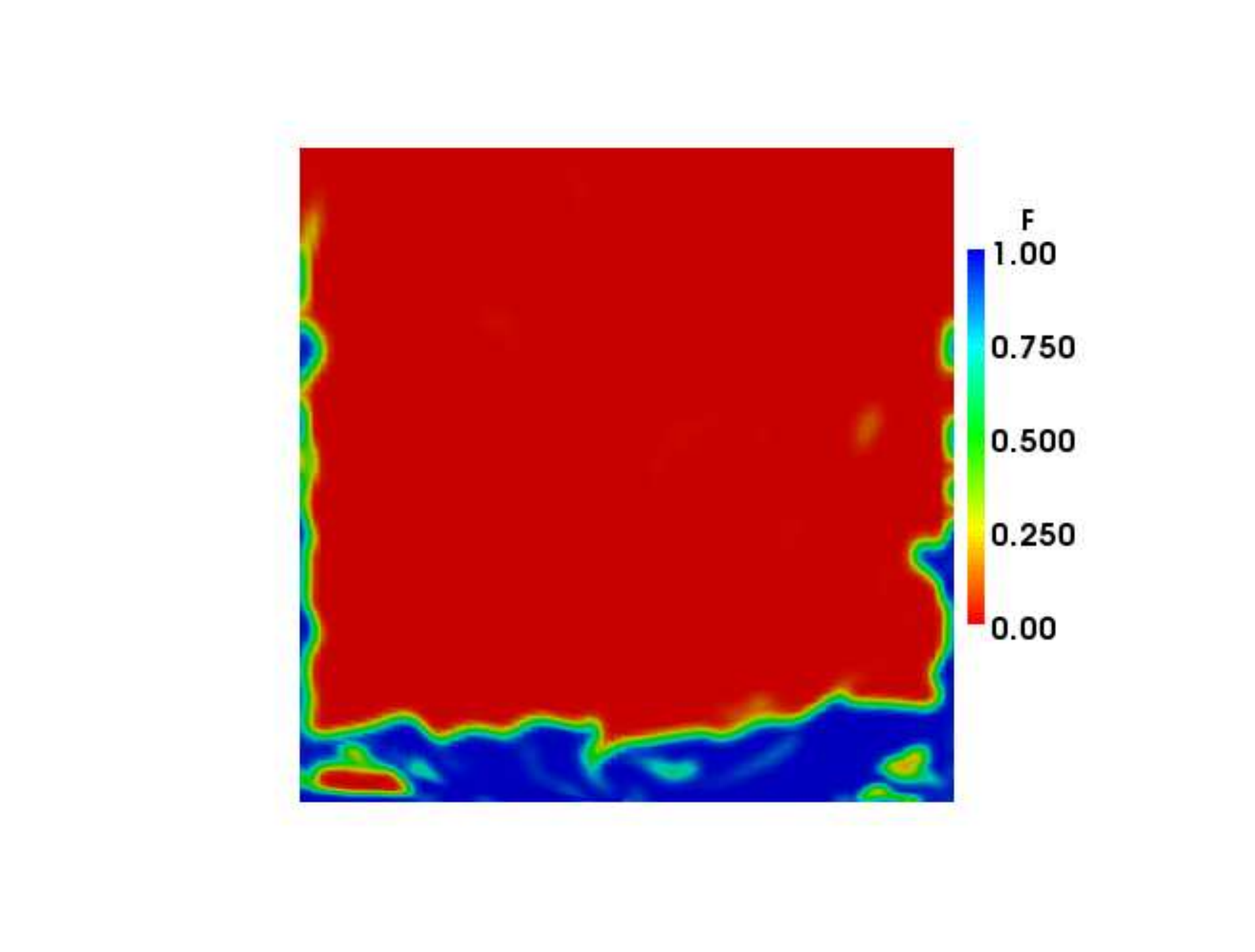}}
\subfigure[$\hspace*{0.1cm} t=2T$]{
\includegraphics[width=.4\textwidth]{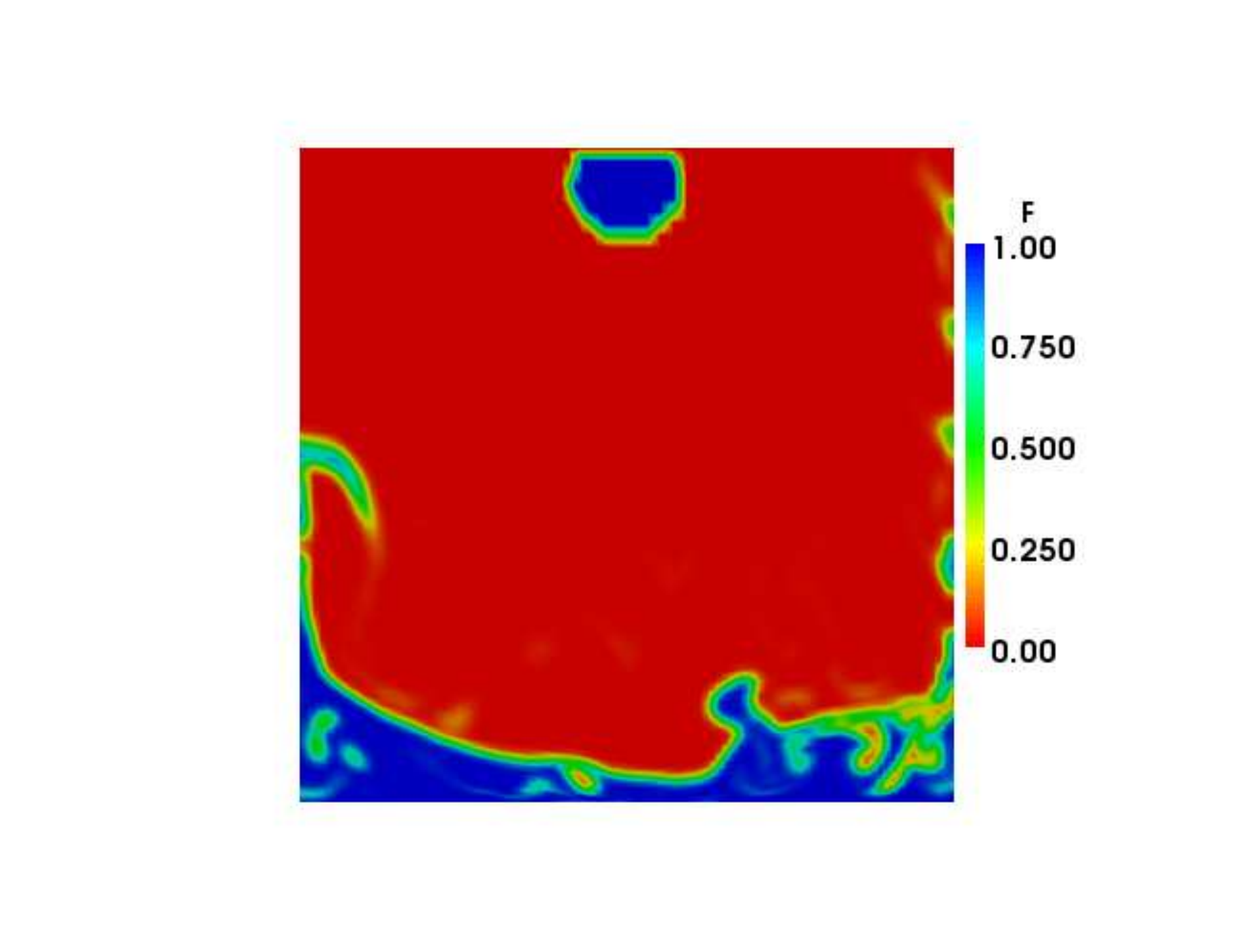}}}
\caption{Snapshots of the indicator function $F(t,\x)$ obtained 
at the indicated instants of time in the solution of the
time--dependent problem \eqref{ns_liquid}--\eqref{eq:vg0} with the
parameters $T=1.0$, $V_d = 1.0$ and $r=0.25$.}
\label{fig:vof}
\end{center}
\end{figure}

Solutions to the problem described above depend on the following three
parameters
\begin{enumerate}
\item  length $T$ of the interval at which droplets are released, 
\item  velocity of the droplet  $V_d$, and
\item  radius of the droplets $r$.
\end{enumerate}
We emphasize that the choice of this particular model problem was in
fact inspired by an industrial application described in Volkov
\emph{et.~al.} \cite{vplg09a} which has also motivated our broader
research program.  Numerical solution of this time--dependent
free--boundary problem is obtained using the solver \texttt{InterFOAM}
which is a part of the library \texttt{OpenFOAM} \cite{openfoam}and is
based on the VoF method. Details of the numerical method and its
implementation in \texttt{InterFOAM} can be found for instance in the
Ph.D. thesis of Rusche\cite{henrik}. The resolution used in our
calculations was $100 \times 100$ grid points with a nondimensional
time step of 0.05. In order to characterize the time--dependent and
mean fields obtained as solutions to this problem, in Figure
\ref{fig:vof} we present several snapshots of the indicator function
$F(t,\x)$ at different time levels spanning two periods of droplet
impingement. To fix attention, the results presented in Figure
\ref{fig:vof} were obtained using the following parameters $T=1.0$,
$V_d = 1.0$ and $r=0.25$.


\section{Algebraic Closure Model}
\label{sec:efs_closure}

\begin{figure}[t]
\centering
\includegraphics[width=0.8 \textwidth]{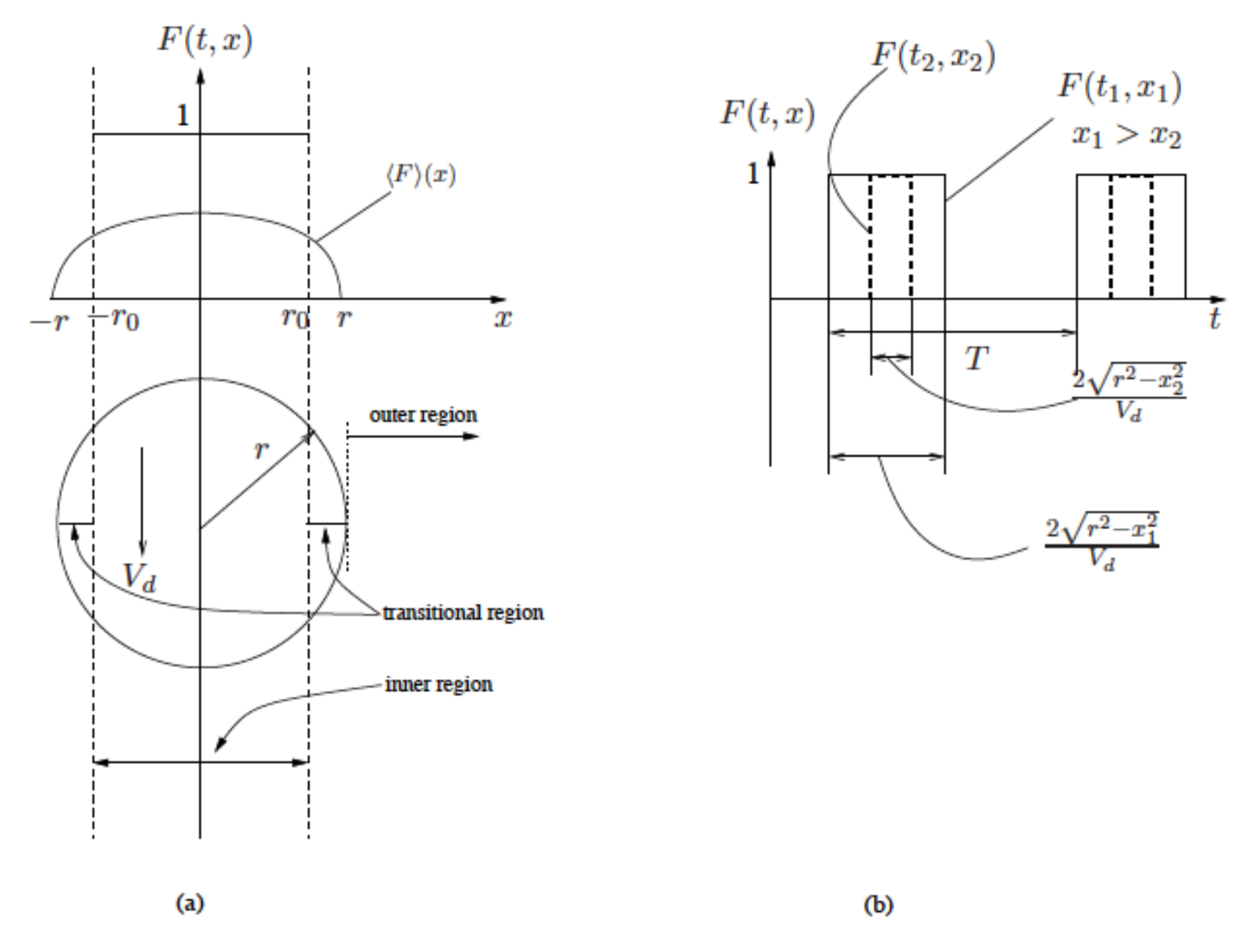} 
\caption{Sketch illustrating the main features of the model based 
on Assumptions \ref{assume3}: (a) construction of the
piecewise--constant approximation $\tF$ to $\langle F \rangle$, (b)
time--dependence of the indicator function $F(x,t)$ for different
values of the distance $x$.}
\label{figavg}
\end{figure}

In Section \ref{sec:efs_avr} we introduced the Reynolds decomposition
of the flow variables into the time--averaged quantities (denoted with
angle brackets $\langle \cdot \rangle$) and fluctuating quantities
(denoted with primes). The terms involving averaged products of
fluctuating quantities appear as unknowns in averaged equations
\eqref{nsR} and must be closed with suitable ``closure models'',
analogous to those which arise in classical turbulence modeling
approaches. The most commonly used methods of turbulence modeling are
surveyed in monograph by Pope\cite{pope}. Briefly speaking, depending
on their mathematical structure, such approaches fall into two main
categories, namely, algebraic models and differential models in which
evolution of the quantities introduced to close the system is governed
by additional PDEs. Some attempts at deriving closure models for
two--phase flows were already made by Brocchini \& Peregrine
\cite{bp01b} and by Brocchini\cite{b02} who obtained such models for
regimes characterized by different values of the turbulent kinetic
energy and the turbulent length scale.  In this Section, we make an
attempt to derive an extremely simple algebraic closure relationship
based on an elementary model of the process defined by the following
set of assumptions (see also Figure \ref{figavg}).
\begin{assume} 
\begin{enumerate}
\item Droplets are spherical with radius $r$ and move as rigid objects,

\item there is no collision or coalescence of droplets,

\item droplets are falling periodically with frequency $T^{-1}$ and 
constant velocity $V_d$,

\item the fluid outside droplets (i.e., the gas phase) is motionless,

\item the mean fields do not depend on the vertical coordinate.
\end{enumerate}
\label{assume3}
\end{assume}
\noindent We observe that Assumption \ref{assume3}b constrains the
problem parameters so that $2 r < V_d \, T$. It is also to be noted
that Assumption \ref{assume3}e implies that the model is effectively
one--dimensional with variations only in the direction normal to the
effective surface. While the above assumptions are rather
far--reaching (in particular, the model does not include any effects
of droplet impingement on the free surface), our objective here is to
provide some preliminary insights concerning computation of effective
free surfaces, and development of closures based on more accurate
models is left to future research (some possible directions are
discussed briefly in Section \ref{sec:final}). We thus proceed to use
Assumptions \ref{assume3} in order to derive expressions for the
fluctuating fields $F'$, $u'$ and $v'$ which will be given in terms of
the mean fields $\langle F \rangle$ (or $\tF$), $\langle u \rangle$
and $\langle v \rangle$. These expressions will be in turn used to
determine the fields $a$ and $\b$ in \eqref{nsR2e}--\eqref{nsR2f}.

The coordinate system is shown in Figure \ref{figavg}. We begin by
observing that in the model problem considered the horizontal velocity
component vanishes identically, i.e., $u(t,x)=0$, $t \ge 0$, $x \in
\RR$, as do its mean and the corresponding fluctuation fields
\begin{equation}
 \langle u \rangle(x) = 0, \quad u'(t,x) = 0, \quad t \ge 0, \ x \in \RR.
\label{eq:v}
\end{equation}
Since the model considered assumes periodic behavior, without loss of
generality we are going to focus on a single period of droplet
impingement, i.e., $t \in [0,T]$, and we also remark that the vertical
velocity component $v$ and the indicator function $F$ are
piecewise--constant functions of time at every point in space. We thus
define the following ``pulse'' function
\begin{equation}\label{pulse}
\prod_\theta (t) =\left \{ \begin{array}{ll}
1, & \quad 0 \leq t \leq \theta \\
0, & \quad \textrm{otherwise}
\end{array} \right.,
\end{equation}
where $0 \le \theta < T$, which allows us to write the following
expression for the indicator function $F$ as a function of time and
the coordinate $x$ (for simplicity, we omit the $y$--dependent phase
shift in this expression, as it does not affect the averages which we
will ultimately compute)
\begin{equation}
F(t,x) = \left\{ 
\begin{alignedat}{2}
&\Pi_{\frac{2 \sqrt{r^2 - x^2}}{V_d}}(t), \quad && 0 \le |x| < r \\
& 0, &&  |x| \ge r
\end{alignedat}\right.,
\label{eq:Ft}
\end{equation}
and then the vertical velocity component becomes $v(t,x) = V_d \,
F(t,x)$ for $t \ge 0$, $x \in \RR$. Next, computing the average over
one period we obtain
\begin{equation}
\langle F\rangle (x)  = \frac{1}{T} \int_0^T F(t,x)\, dt =
\left\{ 
\begin{alignedat}{2}
&\frac{2 \sqrt{r^2 - x^2}}{V_d \, T}, \quad && 0 \le |x| < r \\
& 0, &&  |x| \ge r
\end{alignedat}\right.,
\label{eq:Fm}
\end{equation}
and also have $\langle v \rangle (x) = V_d \, \langle F \rangle
(x)$. These expressions allow us to evaluate the fluctuating fields as
follows
\begin{alignat}{2}
& v'(t,x) = v(t,x) - \langle v \rangle (x) & = & \left\{
\begin{alignedat}{2}
&V_d \, {\Pi}_{\frac{2 \sqrt{r^2 - x^2}}{V_d}}(t) 
- \frac{2 \sqrt{r^2 - x^2}}{T}, \quad && 0 \le |x| < r \\
& 0, &&  |x| \ge r
\end{alignedat}\right., 
\label{eq:v'} \\
& F'(t,x) = F(t,x) - \langle F \rangle (x) & = & \left\{
\begin{alignedat}{2}
& \Pi_{\frac{2 \sqrt{r^2 - x^2}}{V_d}}(t) 
- \frac{2 \sqrt{r^2 - x^2}}{V_d \, T}, \quad && 0 \le |x| < r \\
& 0, &&  |x| \ge r
\end{alignedat}\right..
\label{eq:F'}
\end{alignat}
We note that for $-r \le x \le r$ the averaged indicator function
$\langle F \rangle (x)$ has a quadratic distribution which can be
interpreted as resulting in a {\em smeared} interface. However, in
view of Assumption \ref{assume2}, we require a sharp interface $\tg$
corresponding to a {\em piecewise--constant} indicator function $\tF$
\begin{equation}\label{Fm2}
\tF =\left \{ \begin{array}{ll}
1 & \quad \mid x \mid \, \leq r_0\\
0 & \quad  \mid x \mid \, > r_0
\end{array} \right.,
\end{equation}
\noindent where $r_0$ is the new interface location which can be 
determined based on the principle of mass conservation. It is
expressed using the original smeared \eqref{eq:Fm} and the new
piecewise--constant \eqref{Fm2} indicator functions as follows
\begin{equation}\label{Fmass}
\int_0^{\infty} \langle F \rangle(x)\, dx  =
\int_0^{\infty} \tF\, dx
\end{equation}
from which we obtain 
\begin{equation}\label{r0}
r_0 =  \frac{\pi r^2}{2V_dT}.
\end{equation}
In view of Assumption \ref{assume3}b, it is evident that $r_0
<r$. Therefore, one can recalculate the fluctuating indicator
function, now with respect to $\tF$ given by \eqref{Fm2} and
\eqref{r0}, to obtain
\begin{equation}\label{F'}
F'(t,x) =\left \{ \begin{array}{ll}
\Pi_{\frac{2 \sqrt{r^2 - x^2}}{V_d}}(t) - 1 & \quad \mid x \mid < r_0,\\
\Pi_{\frac{2 \sqrt{r^2 - x^2}}{V_d}}(t)     & \quad r_0  \le \mid x \mid \le r, \\
0 &  \quad \mid x \mid > r. 
\end{array} \right.
\end{equation}
We remark that the obtained expressions \eqref{eq:v'} and
\eqref{eq:F'} for the fluctuating quantities depend only on the model
parameters $\{T, r, V_d\}$ and the position of the effective (sharp)
interface $\tg$. We are thus in the position to calculate the averaged
products of fluctuations appearing as components of tensors $\bA$ and
$\bB$ in \eqref{eq:A} and \eqref{eq:B}. First, we observe that in view
of \eqref{eq:v} we have
\begin{equation}
\langle F' u' \rangle = \langle F' u' u' \rangle = \langle F' u' v' \rangle = 0
\label{eq:F'u'}
\end{equation}
everywhere in the domain. As regards the products of fluctuations
which do not include $u'$, we observe that the form of the expression
will depend on the coordinate $x$, and the following three regions are
distinguished
\begin{enumerate}
\item 
{\em inner} region defined by $\mid x \mid < r_0$, i.e., where
$\tF(x) = 1$ and $\langle F \rangle (x) > 0$,

\item 
{\em transitional} region defined by $r_0 < \mid x \mid < r$, i.e.,
where $\tF(x) = 0$ and $\langle F \rangle (x) > 0$,

\item 
{\em outer} region defined by $\mid x \mid > r$, i.e.,
where $\tF(x) = \langle F \rangle (x) = 0$.
\end{enumerate}
The expressions for the averaged products of fluctuating quantities in
these regions are discussed in three subsections below, whereas in the
last subsection we demonstrate how these expressions can be used to
close the terms $a$ and $\b$ in boundary conditions
\eqref{nsR2e}--\eqref{nsR2f}.

\subsection{Inner Region}
\label{sec:efs_inn}

In the inner region $| x | < r_0$, using \eqref{eq:v'} and
\eqref{eq:F'}, we can write after employing some straightforward
properties of pulse function \eqref{pulse}
\begin{equation}\label{f'v'_a}
F'v'= - \frac{2 \sqrt{r^2-x^2}}{T} \Pi_{\frac{2 \sqrt{r^2 -
x^2}}{V_d}}(t)+\frac{2 \sqrt{r^2-x^2}}{T} .
\end{equation}
Time--averaging expression \eqref{f'v'_a} we obtain
\begin{equation} \label{f'v'_inner}
\langle F'v' \rangle = \frac{2\sqrt{r^2-x^2}}{T}- \frac{4 (x^2-r^2)}{V_d \, T^2}.
\end{equation}
Following the same steps as above, we can deduce that 
\begin{equation}\label{f'v'v'_inner}
\langle F'v'v' \rangle = \frac{8 (r^2-x^2)^{3/2}}{V_d\, T^3} - \frac{4 (r^2-x^2)}{T^2}.
\end{equation}

\subsection{Transitional Region}
\label{sec:efs_trans}

In the transitional region $r_0 < |x| < r$, again using
\eqref{eq:v'} and \eqref{eq:F'}, and following the steps involved in
obtaining \eqref{f'v'_inner} and \eqref{f'v'v'_inner}, we can express
$\langle F'v' \rangle$ and $\langle F'v'v' \rangle$ as
\begin{subequations}\label{average_transition}
\begin{align}
\langle F'v' \rangle &=   \frac{2\sqrt{r^2-x^2}}{T}-  \frac{4 (x^2-r^2)}{V_d \, T^2}, \label{f'v'_trans} \\ 
\langle F'v'v' \rangle &=  \frac{2V_d\sqrt{r^2-x^2}}{T}-\frac{8 (r^2-x^2)}{T^2} + \frac{8 (r^2-x^2)^{3/2}}{V_d \, T^3}. \label{f'v'v'_trans}
\end{align}
\end{subequations}

\subsection{Outer Region}
\label{sec:efs_out}

Since in the outer region $|x| > r$ we have $F'(x) =  0$, it follows that
\begin{subequations}\label{avg_outer}
\begin{align}
\langle F'v' \rangle &= 0,  \label{f'v'_outer}\\
\langle F'v'v' \rangle &=0. \label{f'v'v'_outer}
\end{align}
\end{subequations}

\subsection{Closure Terms in the Boundary Conditions on the Effective Surface}
\label{sec:closureBC}

In this Section we derive expressions for the closure terms $a$ and
$\b$ in boundary conditions \eqref{nsR2e}--\eqref{nsR2f} based on the
simple algebraic closure model proposed here. First, since $u' \equiv
0$ and $\langle u \rangle \equiv 0$, we note that
\begin{equation}
\bA = (\rho_L-\rho_G) \begin{bmatrix}
0                      \\
\langle F'v' \rangle   \\
\end{bmatrix}, \qquad
\bB = (\rho_L-\rho_G)
\begin{bmatrix}
0 & 0 \\
0 &  \langle F'v' v'\rangle  \\
\end{bmatrix},
\label{eq:AB0}
\end{equation}
where the form of the nonzero entries depends on the location with
respect to the effective surface $\tg$ (see Sections \ref{sec:efs_inn},
\ref{sec:efs_trans} and \ref{sec:efs_out}). In our model the location
of the effective surface $\tg$ coincides with the boundary between
the inner and transitional regions, cf.~Figure \ref{figavg}a. In view
of \eqref{eq:ab} and \eqref{eq:AB0} we therefore have
\begin{align}
& \begin{aligned}
a &= (\rho_L-\rho_G) \left(n_y  \langle F'v' \rangle\big|_{\textrm{transitional}} - 
n_y  \langle F'v' \rangle\big|_{\textrm{inner}} \right) = 0,
\end{aligned} \label{eq:a2} \\
& \begin{aligned}
\b &= (\rho_L-\rho_G) \left(
\begin{bmatrix}
0    \\ n_y \langle F'v'v' \rangle   \\
\end{bmatrix}\Bigg|_{\textrm{transitional}}
-
\begin{bmatrix}
0    \\ n_y \langle F'v'v' \rangle   \\
\end{bmatrix}\Bigg|_{\textrm{inner}} \right) \\
   &= (\rho_L-\rho_G)
\begin{bmatrix}
0    \\ n_y \left[ \frac{r}{T^2} \sqrt{4 V_d^2T^2-\pi^2 r^2} - \frac{r^2}{V_d^2 T^4} \left( 4 V_d^2T^2-\pi^2 r^2\right) \right]  
\end{bmatrix},
\end{aligned}
\label{eq:b2}
\end{align}
where we also used the relationships derived in Sections
\ref{sec:efs_inn} and \ref{sec:efs_trans} evaluated at $x=r_0$ to
express the fluctuating terms. Relations \eqref{eq:a2} and
\eqref{eq:b2} close our system of averaged equations \eqref{nsR2}. The
conclusion that $a=0$ is consistent with the fact that there is no
mass production at the effective surface. It should be emphasized that
the proposed closure model is quite problem--specific and it is not
obvious whether the same ideas could be used to develop closures for
other flows with effective free surfaces. The closure model derived by
Brocchini \& Peregrine \cite{bp01b} for an analogous flow regime is
more general, and at the same time less explicit, as it is constructed
in terms of an a priori unspecified ``probability function''
describing ejection of droplets with some specific velocity. The model
proposed in this Section can be regarded as assuming a specific form
of this probability function. A numerical approach well suited for the
solution of free--boundary problem \eqref{nsR2} is described in the
next Section, whereas in Section \ref{sec:efs_results} we will analyze
the dependence of the solutions on the parameters $\{V_d, R, T \}$
characterizing the closure model.

\section{Solution of Averaged Equations with Effective Surfaces 
Via a Shape Optimization Approach}
\label{sec:efs_steady}

In Sections \ref{sec:problem} and \ref{sec:efs_closure} we formulated
a set of steady--state PDEs as a simplified time--averaged model of a
fluid problem involving unsteady free boundaries such as, for example,
the system introduced in Section \ref{sec:model} with droplets
impinging on the pool surface, cf.~Figure \ref{fig:model}. In the
proposed formulation, the steady liquid--gas interface is represented
as the effective free surface described mathematically as the
discontinuity of the averaged indicator function $\tF$. Therefore, the
set of governing equations \eqref{nsR2} has the form of a steady
free--boundary problem. Since such problems tend to be hard to solve
numerically, we argue below that a computationally efficient approach
can be developed by formulating this problem in terms of shape
optimization. More specifically, we will frame it as finding an
optimal shape of the interface (i.e., effective free boundary
$\tilde{\Gamma}_{LG}$) such that a cost functional representing the
residual of one of the interface boundary conditions will be minimized
with respect to the position of the interface subject to the
constraints representing the governing (time--averaged) equations and
the remaining boundary conditions. We refer the reader to the
monograph by Neittaanmaki \emph{et al.}\cite{nst06} for a general
discussion of advantages of such an approach, and to papers by Volkov
\& Protas \cite{vp08a} and Volkov \emph{et al.} \cite{vplg09a} for a
discussion of some applications to problems similar to the one studied
here which also included treatment of contact lines.

System \eqref{nsR2} represents a steady--state free--boundary
problem where $\tilde{\Gamma}_{LG}$ has to be found as a part of the
solution. By fixing the domain and its boundary $\tg$, and removing
one of the boundary conditions, for example, the normal component of
\eqref{nsR2f}, we obtain a steady {\em fixed--boundary} problem. The
residual of the normal component of condition \eqref{nsR2f} is then
minimized with respect to the unknown shape of the effective surface
$\tg$ using a suitable shape--optimization algorithm. The choice of
the normal component of condition \eqref{nsR2f} as the optimization
criterion is motivated by the fact that this condition contains
closure terms, hence in practical situations need not be satisfied up
to the machine accuracy.

\begin{figure}
\centering
\includegraphics[width=0.6 \textwidth]{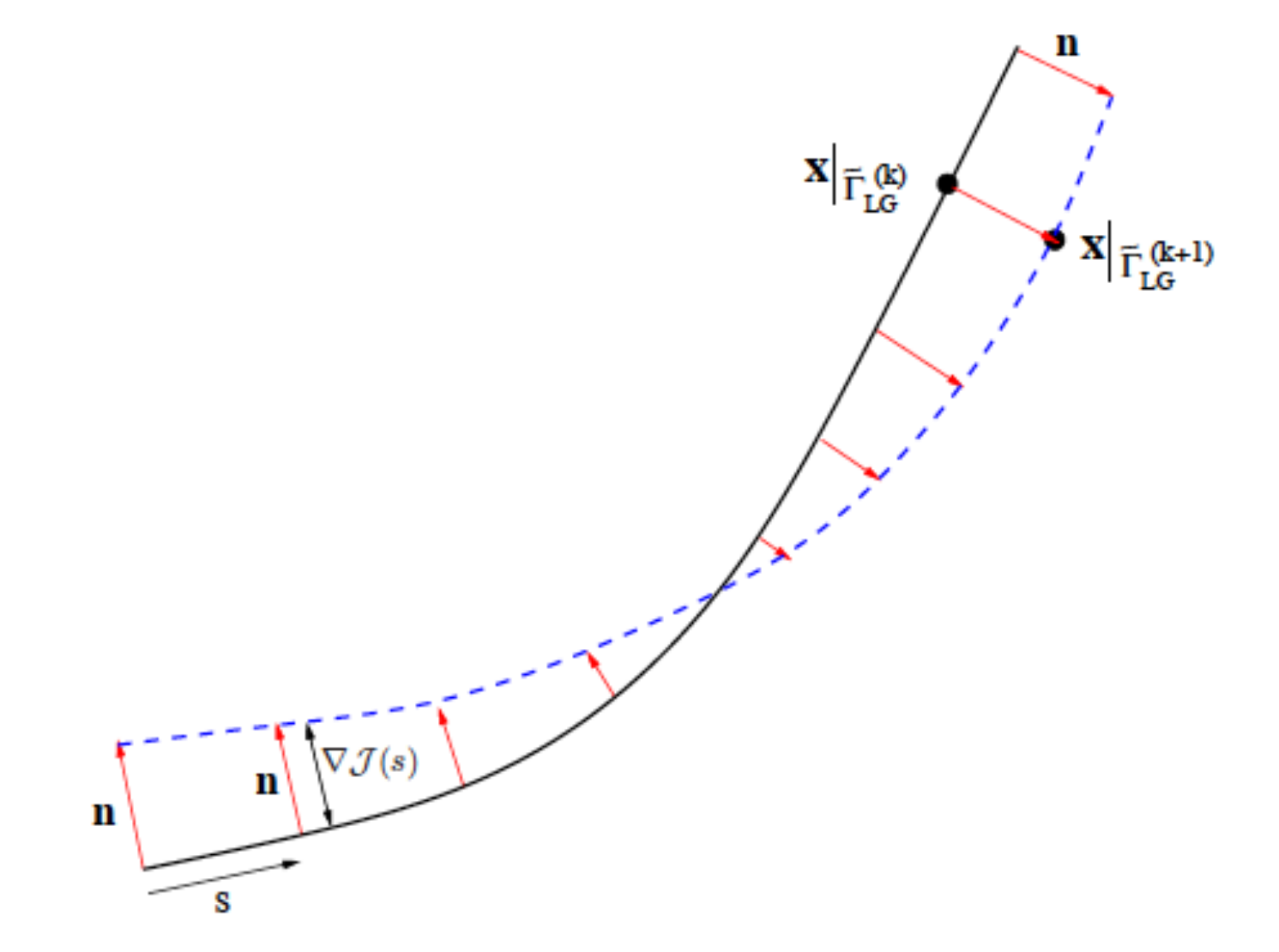} 
\caption{Schematic of a single step of the shape optimization
  algorithm \eqref{eq:iter}: (solid line) approximation to the
  effective surface at the $k$--th iteration and (dashed line)
  approximation to the effective surface at the $k+1$--th iteration.}
\label{fig:gradJ}
\end{figure}
Our solution approach is then formulated as follows. Suppose we define
the function $\chi \; : \; \tg \rightarrow \RR$ which will represent
the residual of the normal component of the momentum boundary
condition \eqref{nsR2f}
\begin{equation}\label{chi}
\chi \triangleq \n \cdot \asigL \cdot \n  - \n \cdot \asigG \cdot \n    - \gamma \kappa\, -\mathbf{b} \cdot \n.
\end{equation}
The cost functional can then be defined as
\begin{equation}\label{J}
\mathcal{J}(\tg)= \frac{1}{2} \int_{\tg}\chi^2 \, d\sigma,
\end{equation}
so that the  optimization problem becomes
\begin{equation}\label{opt_J}
\min_{\tg} \mathcal{J}(\tg),
\end{equation}
where $\chi$ in \eqref{chi} depends on the shape of the effective free
surface $\tilde{\Gamma}_{LG}$, which is the control variable in
optimization problem \eqref{opt_J}, via governing PDEs
\eqref{nsR2a}--\eqref{nsR2d} subject to boundary conditions
\eqref{nsR2e}, \eqref{nsR2g}, and the tangential component of
condition \eqref{nsR2f}, i.e.,
\begin{equation}\label{nsR2h}
\mathbf{t} \cdot \Big[ \langle \sig  \rangle \Big]_L^G \cdot \n =0, \quad  
\textrm{on}  \quad \tilde{\Gamma}_{LG}.
\end{equation}
Thus, we observe that the system of PDEs serving as the constraint for
optimization problem \eqref{opt_J} has in fact the form of a
fixed--boundary problem which makes evaluation of the cost functional
at every iteration easier (i.e., it corresponds to the approximation
$\tg^{\,(k)}$ of the effective surface at the given $k$--th
iteration). The position of the effective interface
$\tilde{\Gamma}_{LG}$ can then be found using the following iterative
gradient--descent algorithm
\begin{equation}
  \x|_{\tg^{\,(k+1)}} = \x|_{\tg^{\,(k)}} +
  \tau_k \,\bG\left[\nabla \J\left(\tg^{\,(k)}\right) \, \n\right],
  \qquad k=1,2,\dots, 
\label{eq:iter}
\end{equation}
where $\x|_{\tg^{\,(k)}} \in \RR^2$ represents points on the interface
$\tg$ at the $k$--th iteration and $\tau_k$ is the length of the step
in the descent direction. The function $\bG$ determines the specific
form of the optimization algorithm used (e.g., the steepest descent,
conjugate gradients, or quasi--Newton method, etc., see
Ref.~\onlinecite{nw00}). In our results reported in Section
\ref{sec:efs_results} we use the Conjugate Gradients Method.  A
central element of algorithm \eqref{eq:iter} is the cost functional
gradient $\nabla \J \, : \, \tg^{(k)} \rightarrow \RR$ representing
the continuous sensitivity of cost functional \eqref{J} to
infinitesimal modifications in the normal direction of the
shape of $\tilde{\Gamma}^{(k)}_{LG}$. In other words, as indicated in
Figure \ref{fig:gradJ}, the scalar--valued function $\nabla
\J\left(\tg^{(k)} \right)$, depending on the arclength coordinate
along the interface, represents the normal displacement of the current
approximation to the effective surface $\tilde{\Gamma}^{(k)}_{LG}$
resulting in the largest possible decrease of cost functional
\eqref{J}, see also Appendix \ref{sec:efs_dJ}. The contact points,
where the effective surface $\tg$ meets the solid boundary $\Gamma_0$,
require special attention and we follow here the approach developed by
Volkov and Protas\cite{vp08a} to deal with the contact line problems
in the context of shape optimization.  Determination of the gradient
$\nabla \J$ requires the solution of a suitably--defined adjoint
system and details concerning its derivation are deferred to Appendix
\ref{sec:efs_grad}. The step size $\tau_k$ is obtained via solution of
the following line minimization problem
\begin{equation}
\tau_k = \argmin_{\tau > 0} \J\left( \x|_{\tg^{\,(k)}} +\tau \,\bG\left[\nabla \J\left(\tg^{\,(k)}\right) \, \n\right]\right)
\label{eq:tauk}
\end{equation}
which can be done, for example, using Brent's method, see
Ref.~\onlinecite{nw00}. The complete approach is summarized in
Algorithm 1.
\begin{table}
\rule{\textwidth}{0.05cm}
{ALGORITHM 1: Iterative minimization algorithm for solving system \eqref{nsR2} via a shape optimization approach. 
\vspace*{-0.4cm}
\begin{flushleft}
\textbf{Input:} $\epsilon_{\tau}$ (adjustable tolerance), 
$\tg^{(0)}$ (initial guess for the effective surface) \hfill \\ 
\textbf{Output:} $\tg$ (effective surface) \hfill
\end{flushleft}} 
\vspace*{-0.5cm}
\rule{\textwidth}{0.025cm}
\vspace*{-0.75cm}
\begin{algorithmic}
\STATE $n \leftarrow 0$
\STATE $\tg^{(0)} \leftarrow$ initial guess (Figure \ref{fig:geom})
\REPEAT
 \STATE solve direct system
 \STATE solve adjoint system
 \STATE compute gradient $\nabla \mathcal{J} \left(\tg^{\,(n)}\right)$
 \STATE perform line minimization to determine $\tau^{(n)}$
 \STATE update effective surface using \eqref{eq:iter}
 \STATE $n \leftarrow n+1$;
\UNTIL $ \tau^{(n)}  < \epsilon_{\tau}$. 
\end{algorithmic}
\vspace*{-0.25cm}
\rule{\textwidth}{0.025cm}
\end{table}

Our time--dependent model problem \eqref{ns_liquid}--\eqref{eq:vg0} is
set up such that the mass of liquid remains constant over every period
of droplet impingement (the amount of mass drained at the bottom of
the container equals the mass added in the form of droplets at the top
of the domain, cf.~Section \ref{sec:model}).  Therefore, it is
necessary to ensure that in our steady--state averaged problem with
effective surfaces \eqref{nsR2} the same mass is enclosed in the
liquid domain $\tOmega_L$.  Mathematically, this is implemented by
constructing an initial guess for the liquid domain $\tOmega_L^{(0)}$
which has the prescribed mass and then making sure that this mass is
not changed at subsequent iterations. For this to happen, it is
required that the shape gradients $\nabla \mathcal{J}(\tg)$ do not
change the volume of $\tOmega_L$. This property is enforced at each
iteration as follows. First, we calculate the mean value $M \in \RR$
of the gradient on the effective surface
\begin{equation}\label{eqm1}
M = \frac{1}{L} \, \int_0^L  \nabla \mathcal{J} (s) ds
\end{equation}
where $s \in [0,L]$ is the corresponding arclength coordinate.  The
new gradient with zero mean displacement in the normal direction is
then obtained as
\begin{equation}\label{eqm2}
  \widetilde{ \nabla \mathcal{J}}(s) = \nabla \mathcal{J}(s)-M, \qquad \forall_{s \in [0,L]}. 
\end{equation}
The cost functional gradient $\nabla \mathcal{J}$ is replaced with
zero--mean gradient $ \widetilde{\nabla \mathcal{J}}$ in expressions
\eqref{eq:iter} and \eqref{eq:tauk}.

\section{Results and Discussions}
\label{sec:efs_results}

\begin{figure}
\begin{center}
   \includegraphics[width=0.6 \textwidth]{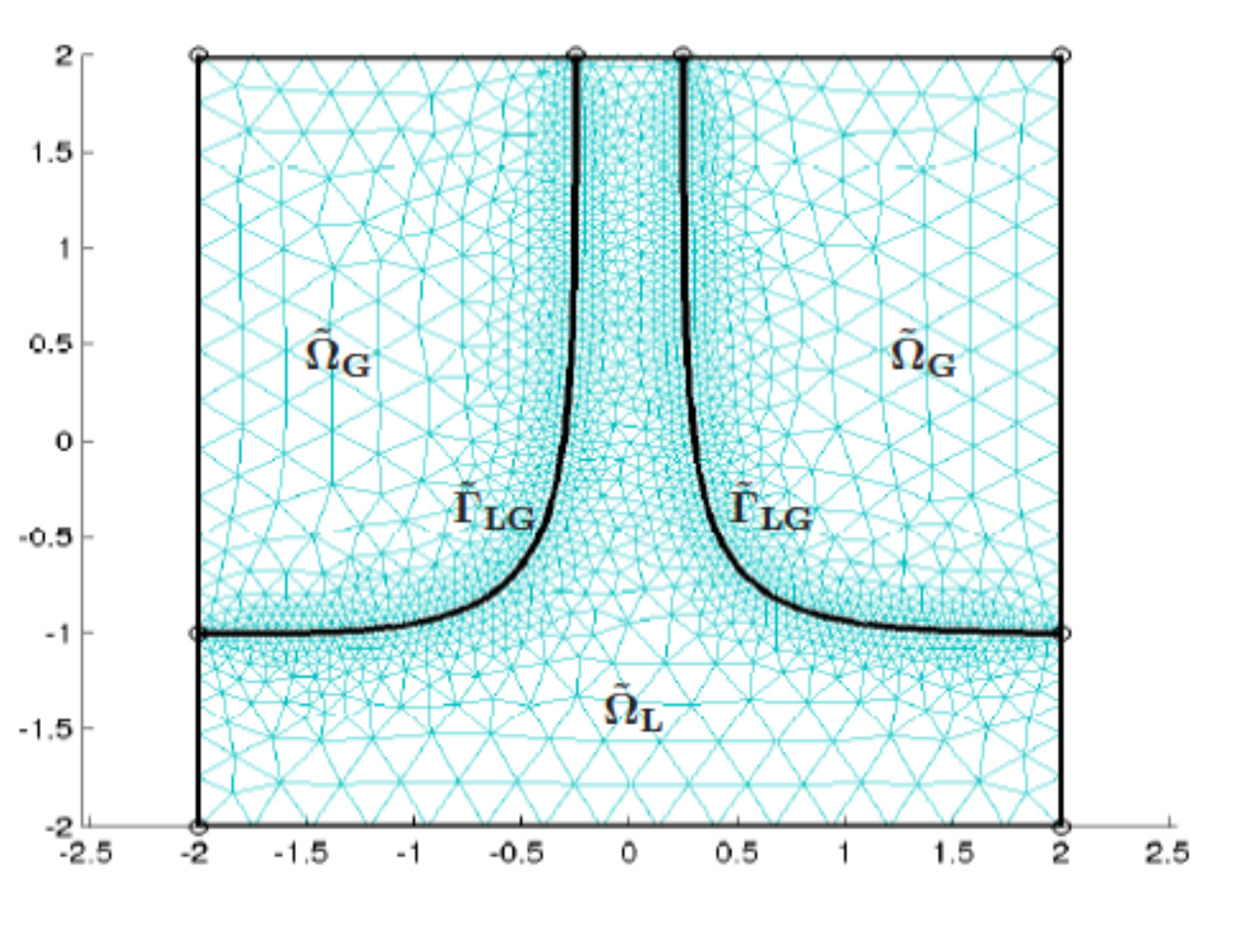}
      \end{center}
\caption{Geometry of the computational domain $\Omega$ with the liquid 
and gas subdomains $\tOmega_L$ and $\tOmega_G$ and the effective
boundary $\tg$ used as the initial guess in shape optimization
algorithm \eqref{eq:iter}. The figure also shows the unstructured
triangular finite--element mesh used in the solution of direct and
adjoint problems.}
\label{fig:geom}
\end{figure}

\begin{figure}
\begin{center}
\mbox{\hspace*{-1.0cm}
\subfigure[  ]{
\label{fig3}
\includegraphics[width=0.55\textwidth]{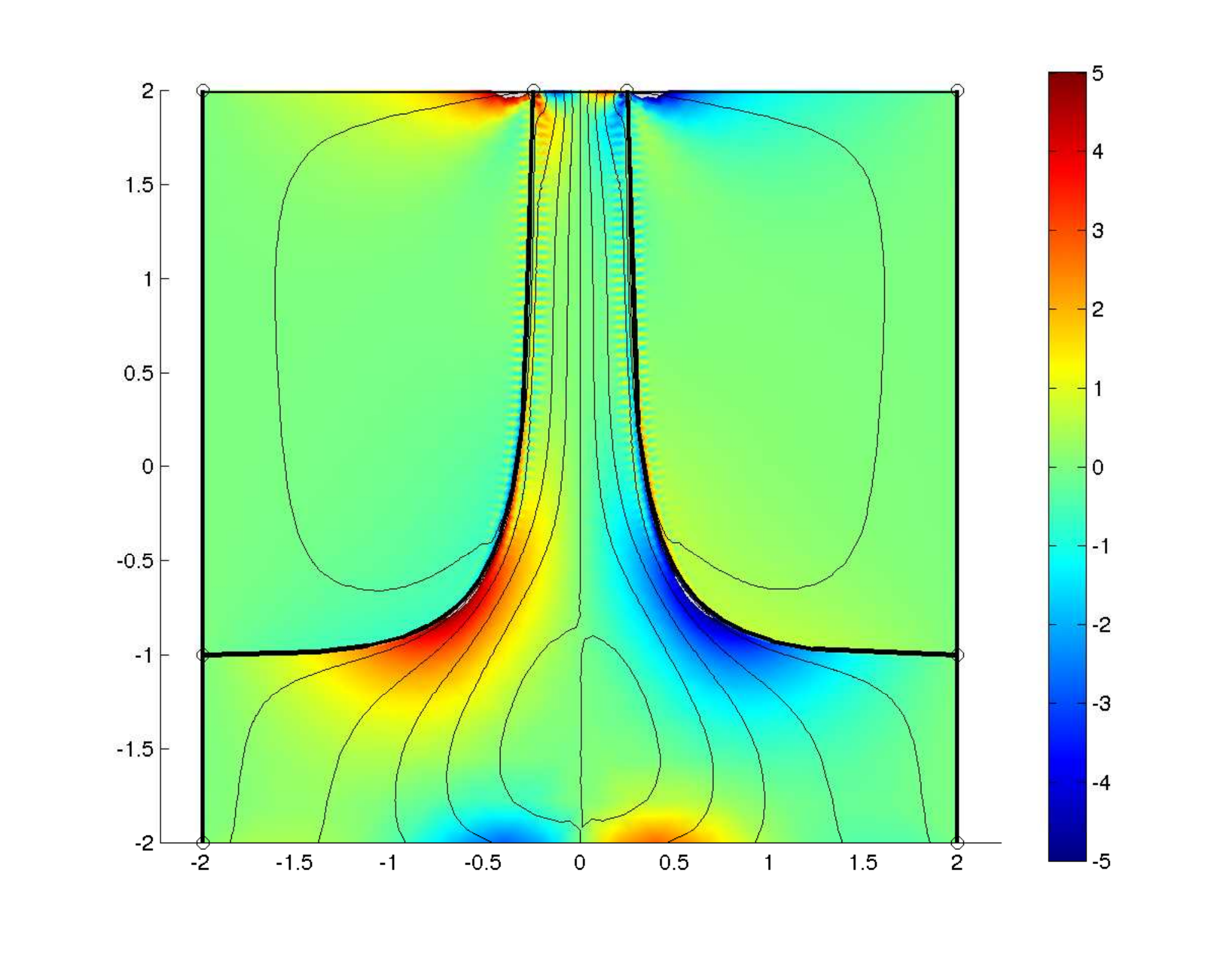}}
\subfigure[]{
\label{fig4}
\includegraphics[width=0.55\textwidth]{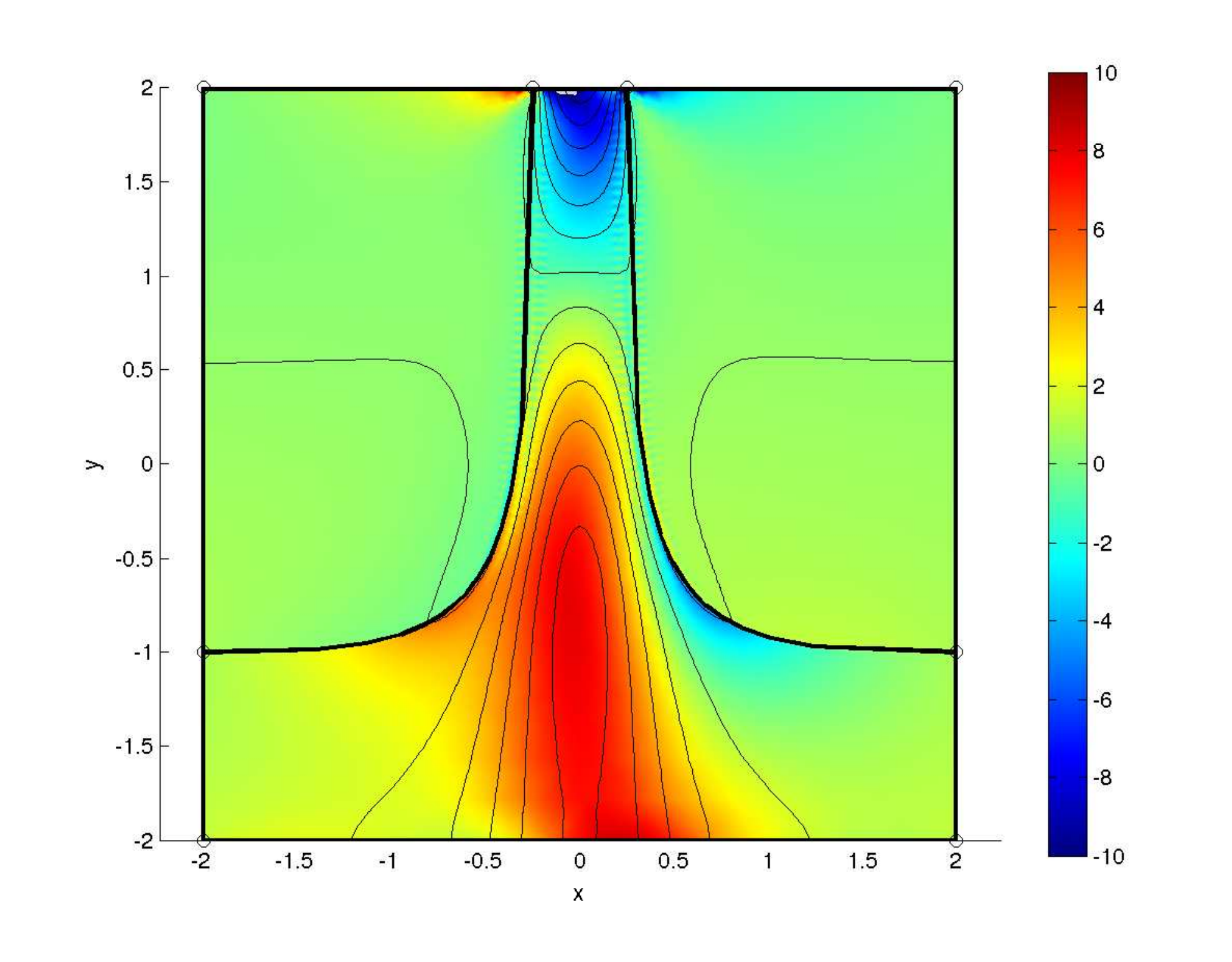}}}
\end{center}
\caption{Vorticity fields and streamlines in the numerical solution 
of (a) direct problem and (b) adjoint problem at the first iteration.}
\label{figsec}
\end{figure}

In this Section we present sample computations for the problem of
determining the effective free surfaces in the flow described in
Section \ref{sec:model}, see also Figure \ref{fig:model}. In order to
calculate the cost functional gradient (given by expression
\eqref{final_gradient} in Appendix \ref{sec:efs_grad}) we need to solve 
``direct'' system \eqref{nsR2} and adjoint system
\eqref{ns_adjoint_liquid}--\eqref{ns_adjoint_gas}. Both these
solutions are obtained using the finite--element method implemented in
the COMSOL script environment. The domain (Figure \ref{fig:geom}) is
discretized using approximately 4000 Lagrangian elements with mesh
size varying between 0.9 to 0.01. In all computations presented here
we used the physical parameters with values indicated in Table
\ref{val} and these calculations were performed using the
Navier--Stokes and Poisson solvers available in COMSOL. For
illustration purposes, in Figures \ref{figsec}a and \ref{figsec}b we
show the fields of the direct and adjoint vorticity obtained at the
first iteration. Before analyzing the solutions with effective free
surfaces obtained for different parameters of the closure models, we
validate the calculation of the cost functional gradient which is the
main element of our computational approach, cf.~Section
\ref{sec:efs_steady} and Appendix
\ref{sec:efs_grad}.

\begin{table}
\begin{centering}
 \begin{tabular}{|c|c|}
\hline
Physical Parameter& Value \\
\hline
Density of the liquid, $\rho_L$ & $1000 [Kg. m^{-3}]$\\

Dynamic viscosity of the liquid, $\mu_L$ & 0.001$ [Pa.s] $\\
Density of the gas, $\rho_G$ & $1[Kg. m^{-3}]$\\

Dynamic viscosity of the gas, $\mu_G$ & 0.00001$ [Pa.s] $\\

Gravitational acceleration, $\g$ & $9.81 [m.s ^{-2}]$\\
Surface tension, $\gamma$ & $0.7 [N. m^{-1}]$\\
\hline
\end{tabular}
\caption{Values of the physical parameters used in the computation.}
\label{val}
\end{centering}
\end{table}


\subsection{Validation of the Shape Optimization Approach}
\label{sec:efs_kappa}

\begin{figure}
\centering
 \includegraphics[width=0.65 \textwidth]{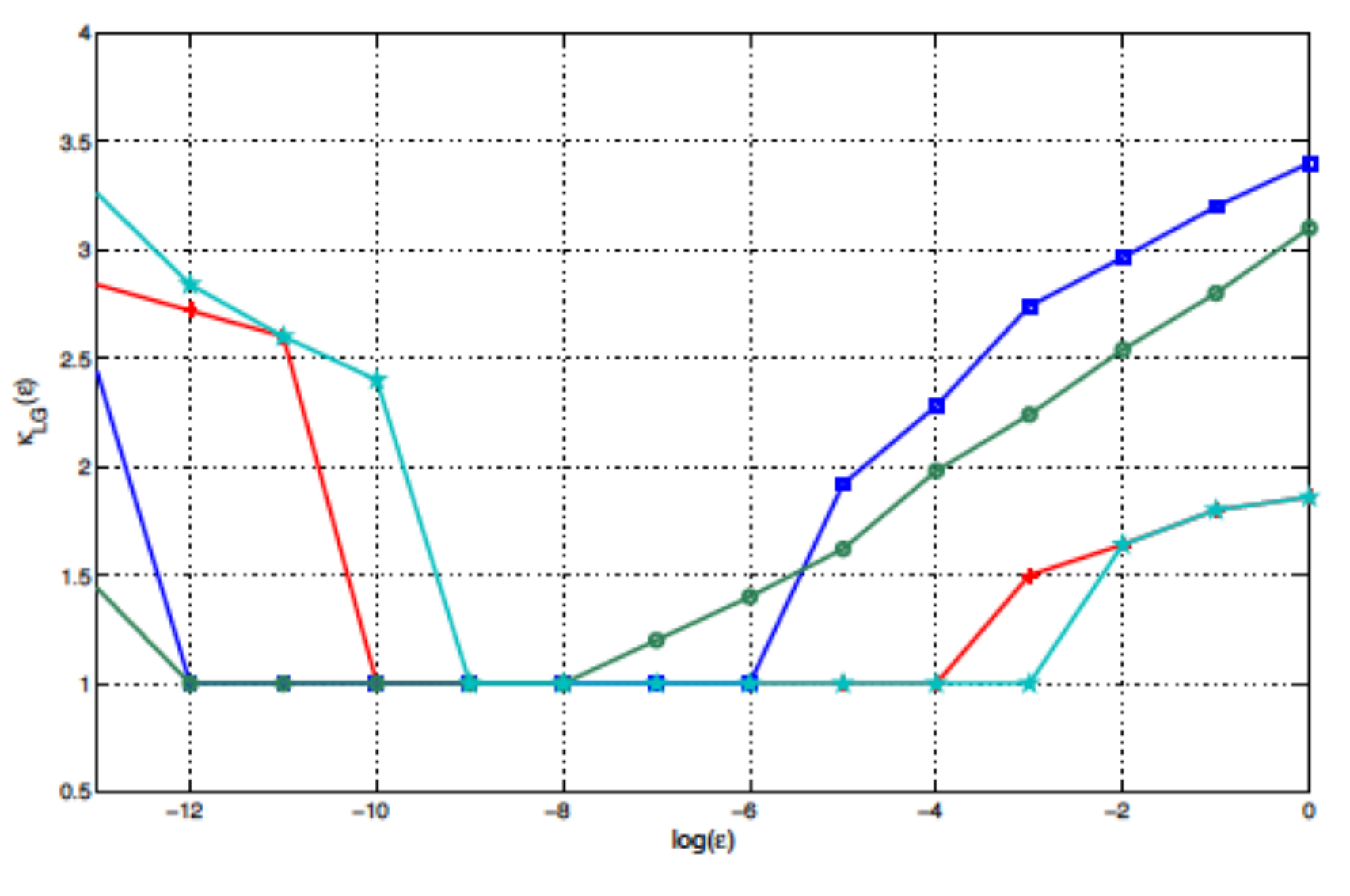}
 \caption{The diagnostic quantity $\varkappa_{\Gamma_{LG}}(\epsilon)$
 as a function of $\epsilon$ for different perturbations ($\star$)
 $\zeta'(s) = \sin(2\pi s / L)$, ($\circ$) $\zeta'(s) = \sin(2\pi s /
 L)$, ($\ast$) $\zeta'(s) = \arcsin(2s/L - 1)$, ($\Box$) $\zeta'(s) =
 \arccos(2s/L - 1)$, where $0\le s \le L$. }
\label{figkappa1}
      \includegraphics[width=0.65 \textwidth]{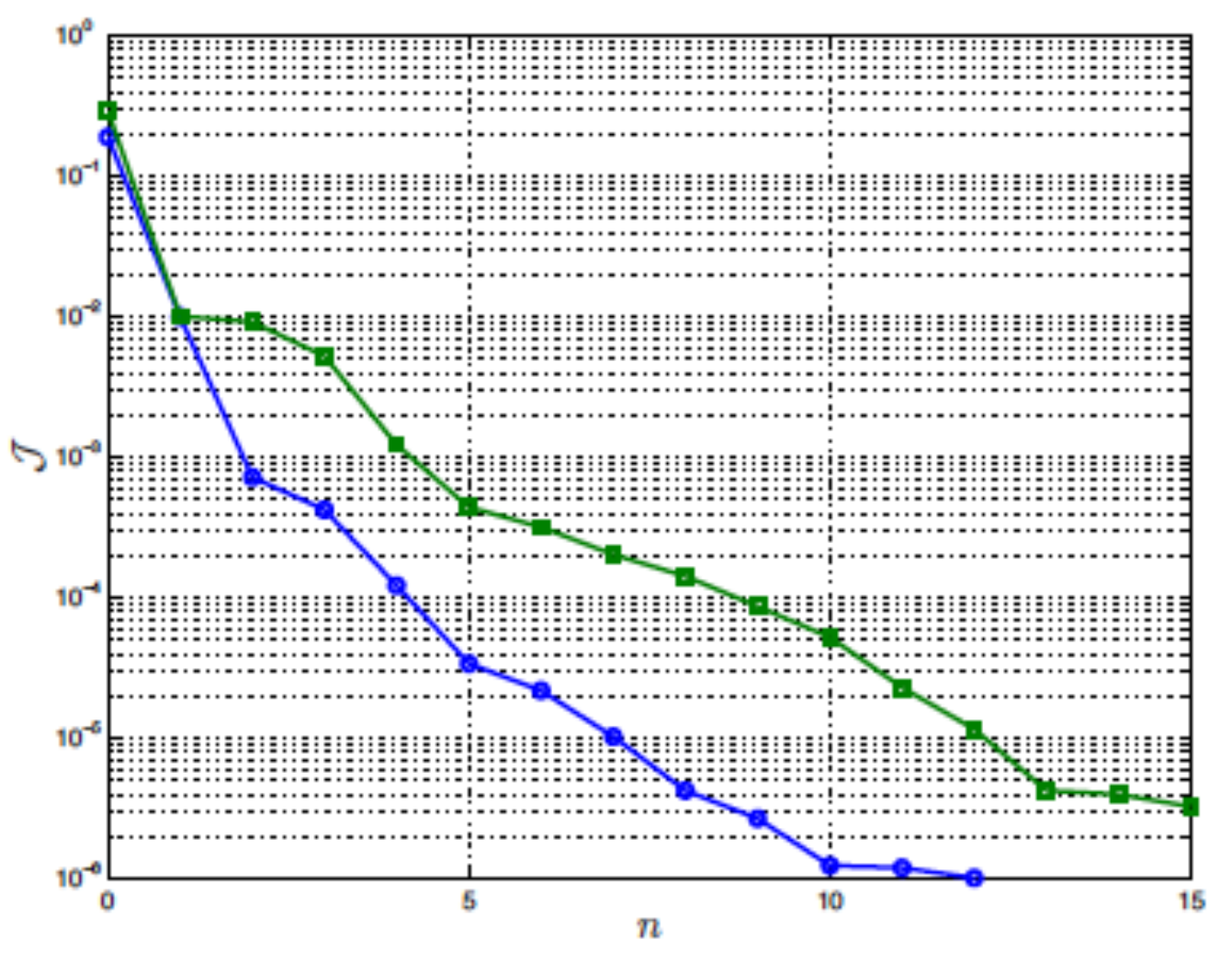}
\caption{Cost functional $\J(\tg^{(n)})$ as a function of the 
iteration count $n$ for ($\bullet$) the case without closure terms and
($\blacksquare$) the case with closure terms $a$ and $\b$ in boundary
conditions \eqref{nsR2e}--\eqref{nsR2f}. The values of the problem
parameters are $V_d= 1.0$, $r=0.25$, and $T=3$. }
\label{Fig:CF}
\end{figure}

In this Section we demonstrate the consistency of the gradients
$\nabla_{\tg}\mathcal{J}$ obtained using expression
\eqref{final_gradient} in Appendix \ref{sec:efs_adj}. A standard test
consists in computing the G\^ateaux differential of cost functional
$\mathcal{J}(\tg)$ in some arbitrary direction $\x' =
\zeta'\n$ using a finite--difference technique and comparing it with
the expressions for the same differential obtained using the gradient
$\nabla_{\tg}\mathcal{J}$ and Riesz representation formula
\eqref{riesz}. The ratio of these two expressions, which is a function
of the finite--difference step size $\epsilon$, is defined as
\begin{equation}\label{kappa_test}
\varkappa_{\tg}(\epsilon)\triangleq
\frac{\mathcal{J}(\x_{\tg}+\epsilon \zeta'
  \n)-\mathcal{J}(\tg)}{\epsilon \big\langle
    \nabla_{\tg}\mathcal{J}, \zeta' \big\rangle}.
\end{equation}
Proximity of $\varkappa_{\tg}(\epsilon)$ to the unity is thus a
measure of the accuracy of the cost functional gradient computed based
on the adjoint field.  Figure \ref{figkappa1} shows the behavior of
the quantity $\varkappa_{\tg}$ as a function of the parameter
$\epsilon$ for different perturbations $\zeta'$. We note that in all
cases the quantity $\varkappa_{\tg}(\epsilon)$ is quite close to unity
for $\epsilon$ spanning over 5 orders of magnitude which indicates
that our gradients are evaluated fairly accurately.  Figure
\ref{figkappa1} reveals deviations of $\varkappa_{\tg}(\epsilon)$ from
unity for large values of $\epsilon$, which is due to truncation
errors, and also for very small $\epsilon$, due to round--off errors,
both of which are well--known effects.  These inaccuracies do not
affect the optimization process, since the deviations observed for
very small $\epsilon$ are only an artifact of how expression
\eqref{kappa_test} is evaluated, whereas large values of $\epsilon$
(or, equivalently, $\tau$) are outside the range of validity of the
linearization on which the optimization approach is based. We also
performed a grid--refinement study of the cost functional gradients
which indicated that the calculation of the gradients is not sensitive
to the resolution. Figure \ref{Fig:CF} shows the decrease of cost
functional \eqref{J} in the case with and without the closure terms
$a$ and $\b$ in boundary conditions \eqref{nsR2e}--\eqref{nsR2f} as a
function of the number of iterations. We observe that the proposed
algorithm results in a steady convergence despite the complicated
nature of the problem, although the rate of convergence is relatively
slow, especially in the case when the closure model is present.

\subsection{Effective Surfaces for Different Parameters in the Closure Model}
\label{sec:efs_efs}

\begin{figure}
\centering
\mbox{\hspace*{-1.0cm}
\subfigure[]{
\Bmp{0.5\textwidth} \hspace*{-1.0cm}
      \includegraphics[width=1.3 \textwidth]{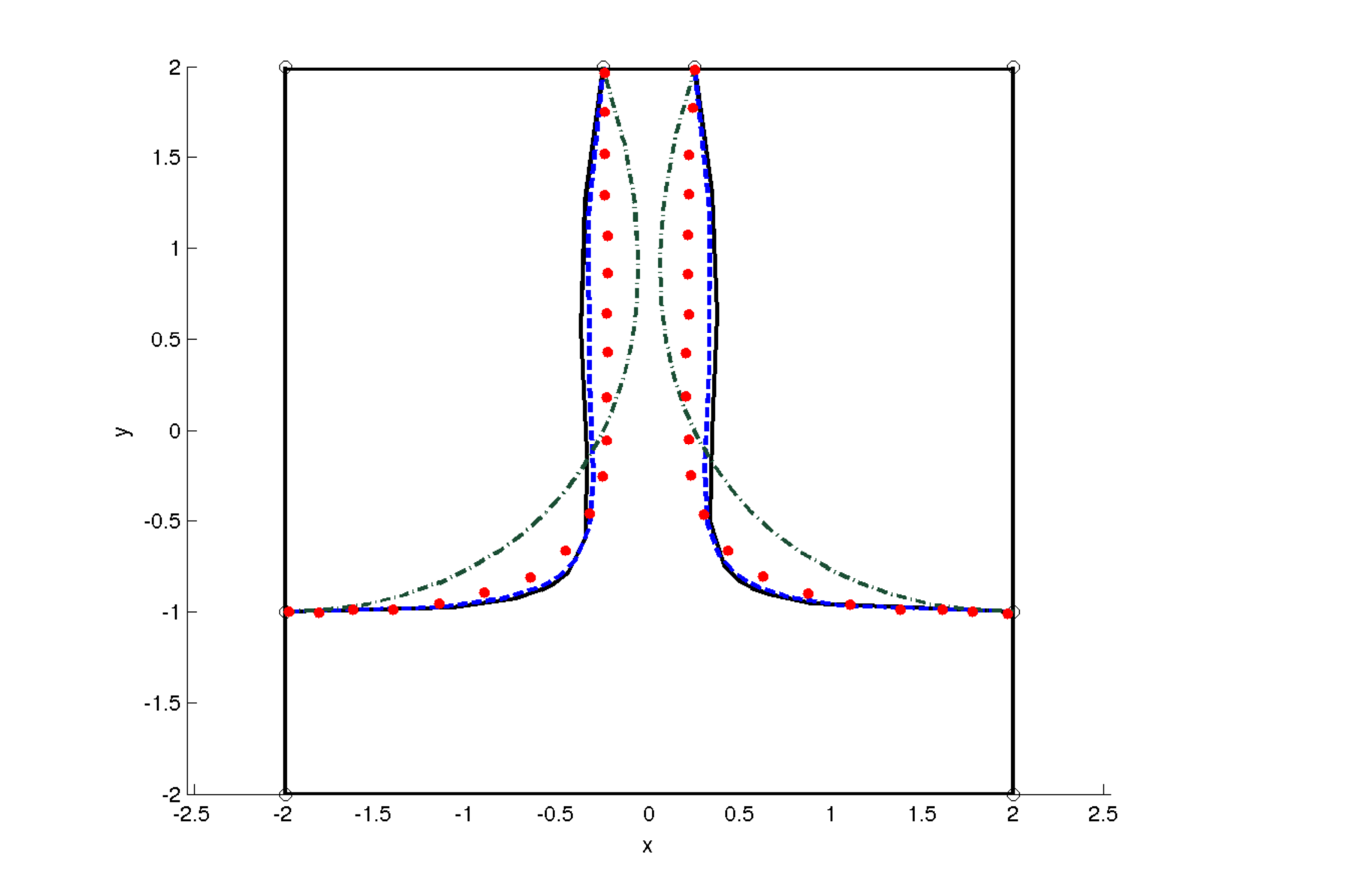}
\centering
 '-----' No closure, ' - - - ' $r=0.1$, \\
'$\cdots$', $r=0.25$, '$- \cdot - \cdot$', $r=0.5$ \\
with $V_d=1.0$ and $T=3$ \\ \vspace*{0.25cm}
\label{Fig:EFSR}
\Emp
}
\subfigure[]{
\Bmp{0.5\textwidth} \hspace*{-1.0cm}
      \includegraphics[width=1.3 \textwidth]{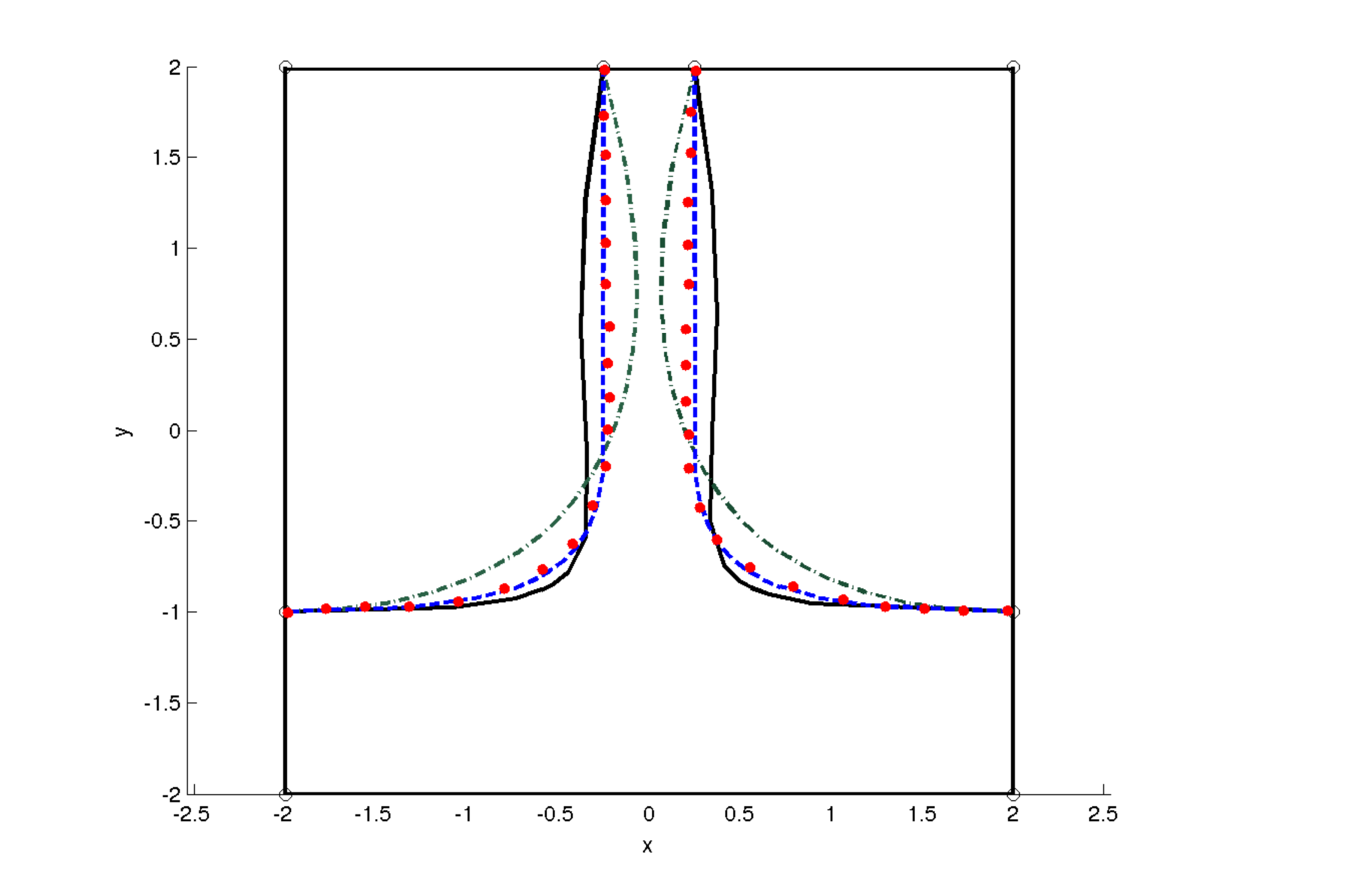}
\centering
'-----' No closure, '$- \cdot - \cdot$', $T=1$, \\
'$\cdots$',  $T=3$, ' - - - ' $T=6$',\\ 
with $V_d=1.0$ and $r=0.25$ \\ \vspace*{0.25cm}
\label{Fig:EFST}
\Emp
}
}
\mbox{\hspace*{-1.0cm}
\subfigure[]{
\Bmp{0.5\textwidth} \hspace*{-1.0cm}
      \includegraphics[width=1.3\textwidth]{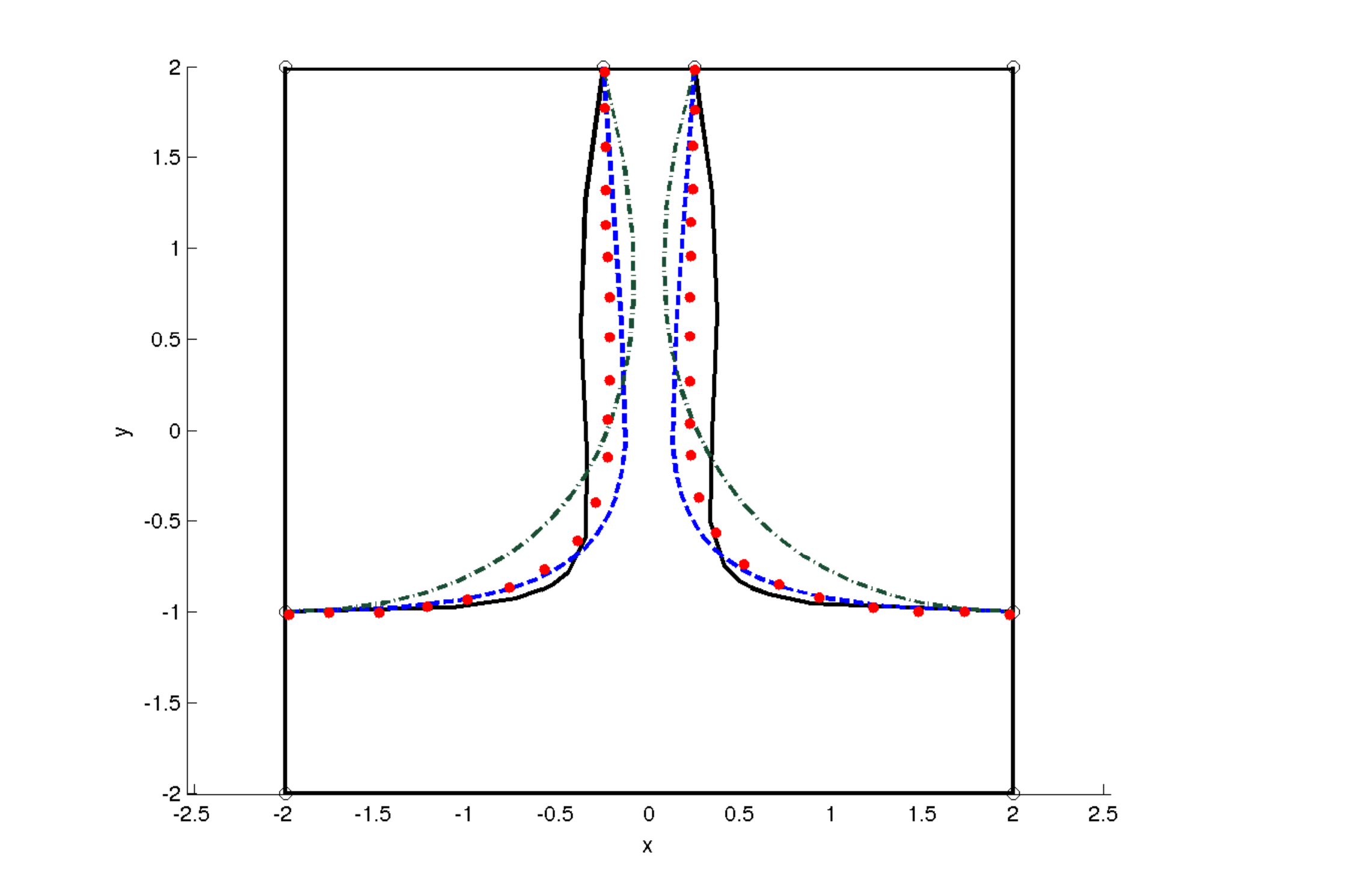} \\
\centering
'-----' No closure,  ' $- \cdot - \cdot$ ' $V_d= 0.5$, \\
 '$\cdots$' $V_d= 1.0$,  ' - - - ' $V_d=1.25$,\\
 with $r=0.25$ and $T=3$ \\ \vspace*{0.25cm}
\label{Fig:EFSV}
\Emp
}
\subfigure[]{
\Bmp{0.5\textwidth} \hspace*{0.3cm}
\includegraphics[width=1.0 \textwidth]{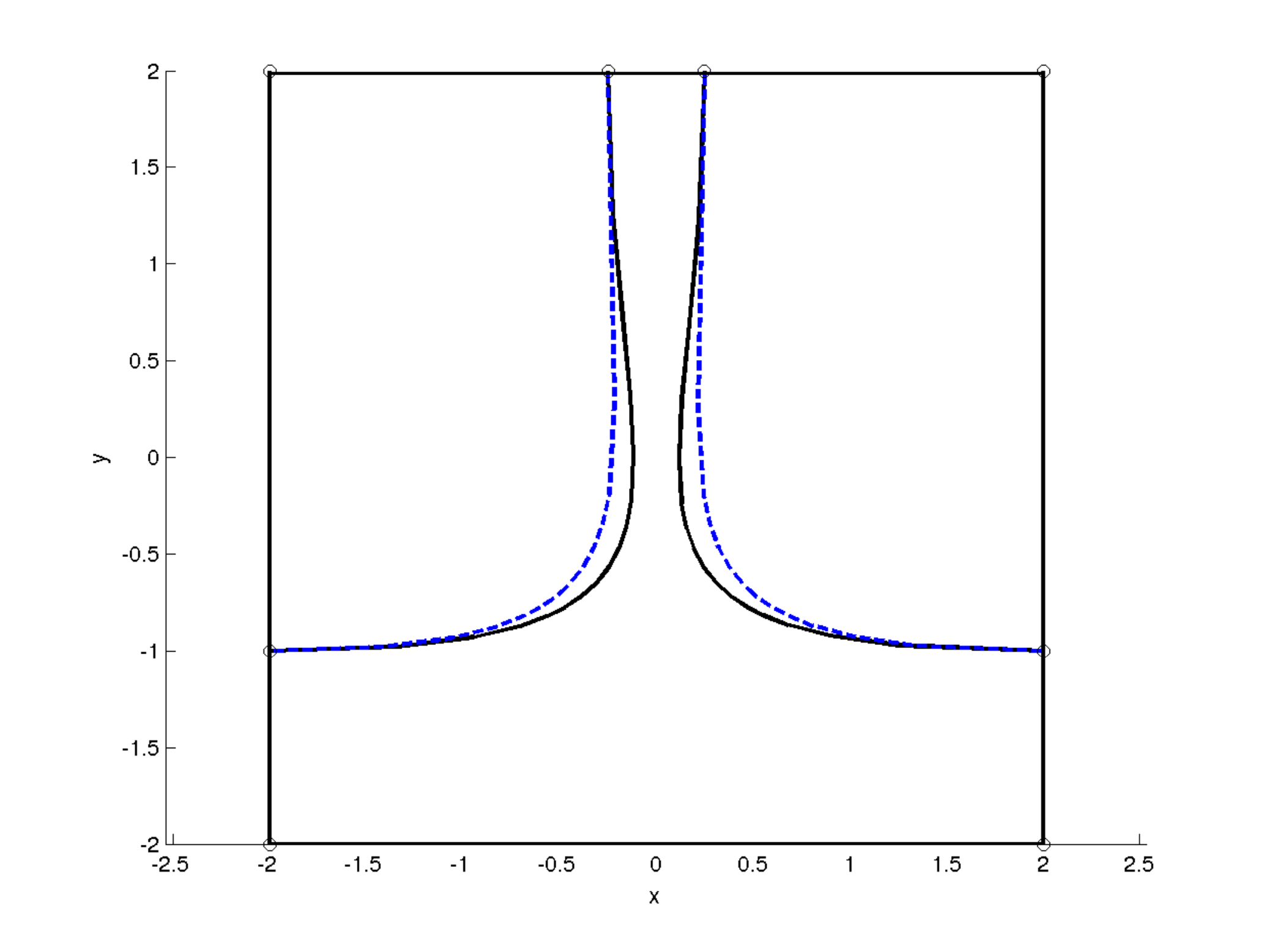}
\centering
'-----' data from the unsteady problem \\
    (see Figure \ref{fig:b}), \\ 
'$ - - -$' algebraic model from Section \ref{sec:efs_closure} \\ \vspace*{0.15cm}
\label{Fig:SFVOF}
\Emp
}}
\caption{Dependence of the shape of the effective surface $\tg$ on the
  parameters of the algebraic closure model introduced in Section
  \ref{sec:efs_closure}: (a) droplet radius $r$, (b) frequency
  $T^{-1}$ of droplet impingement, and (c) droplet velocity $V_d$ with
  other parameters held fixed (see captions of individual subfigures);
  Figure (d) shows the comparison of the effective free surfaces $\tg$
  obtained using the algebraic closure model from Section
  \ref{sec:efs_closure} and the approach described in Section
  \ref{sec:efs_efs} in which terms $a$ and $\b$ are evaluated based on
  the actual data obtained from the time--dependent flow problem with
  the parameter values $V_d=1.0$, $r=0.25$ and $T=3$.}
\label{fig:EF}
\end{figure}

In this Section we employ the computational approach developed in
Section \ref{sec:efs_steady} and validated in Section
\ref{sec:efs_kappa} to construct effective surfaces corresponding to
different values of the three parameters $\{V_d, r, T \}$
characterizing the algebraic closure model introduced in Section
\ref{sec:efs_closure}. In order to reveal different trends, in Figures
\ref{fig:EF}a,b,c we show the effective surfaces obtained by changing
one parameter with the other two held fixed. For comparison, in these
Figures we also include the effective surfaces obtained without any
closure model (i.e., with $\b = \0$ in \eqref{nsR2f}). The parameters
are chosen in such a way that the case with $\{V_d=1.0, r=0.25, T=3
\}$ is present in all three Figures \ref{fig:EF}a,b,c where it
represents the intermediate solution. First of all, we observe that in
all cases smooth effective surfaces have been obtained. As regards the
results shown in Figures \ref{fig:EF}a and \ref{fig:EF}b, we observe
that the effective surfaces approach the effective surface obtained in
the case with no closure as $r \rightarrow 0$ and $T \rightarrow
\infty$, respectively. This is consistent with the fact that $\lim_{r
  \rightarrow 0} \b = \0$ with $V_d$ and $T$ fixed, and $\lim_{T
  \rightarrow \infty} \b = \0$ with $V_d$ and $r$ fixed,
cf.~\eqref{eq:b2}.  In the other limits, i.e., for large $r$ (Figure
\ref{fig:EF}a) and small $T$ (Figure \ref{fig:EF}b), we observe that
the liquid column $\tOmega_L$ becomes much thinner.  As regards the
dependence on droplet velocity $V_d$, from \eqref{eq:b2} we observe
that $V_d = \frac{\pi r}{2 T} < \frac{ 2 r}{T}$ would correspond to
the case with a vanishing closure model (i.e., $\b=\0$), however, this
value of $V_d$ is outside the range of parameter values consistent
with Assumption \ref{assume3}b. Hence, convergence of effective
surfaces to the surface corresponding to the case with no closure is
not observed in Figure \ref{fig:EF}c. We observe that in the proposed
model the closure terms contribute additional flux of momentum in the
direction tangential to the effective surface which can be interpreted
as additional shear stress, cf.~\eqref{nsR2f} and \eqref{eq:b2}. In
the cases with large $r$ and small $T$, corresponding to a thinner
liquid column $\tilde{\Omega}_L$, the effect of the closure model
could be compared to an increase of the surface tension (although this
analogy is rather superficial, since the surface tension contributes
to the normal stresses).  We also add that, in addition to the
effective surfaces presented in Figures \ref{fig:EF}a,b,c, for some
parameter values we also found solutions featuring asymmetric
effective surfaces. This nonuniqueness of solutions is a consequence
of the nonlinearity of the governing system which is reflected in the
nonconvexity of optimization problem \eqref{opt_J}. Since these
asymmetric solutions are not physically relevant, at least not from
the point of view of the actual applications of interest to us, we do
not discuss them in this work.  In problems with multiple solutions in
which such selection cannot be done based on the properties of
symmetry, one can identify the relevant solution as the one
corresponding to the smallest value of cost functional $\J(\tg)$
reflecting the smallest residual \eqref{chi}.

\begin{figure}
\begin{center}
\includegraphics[width=1.0\textwidth]{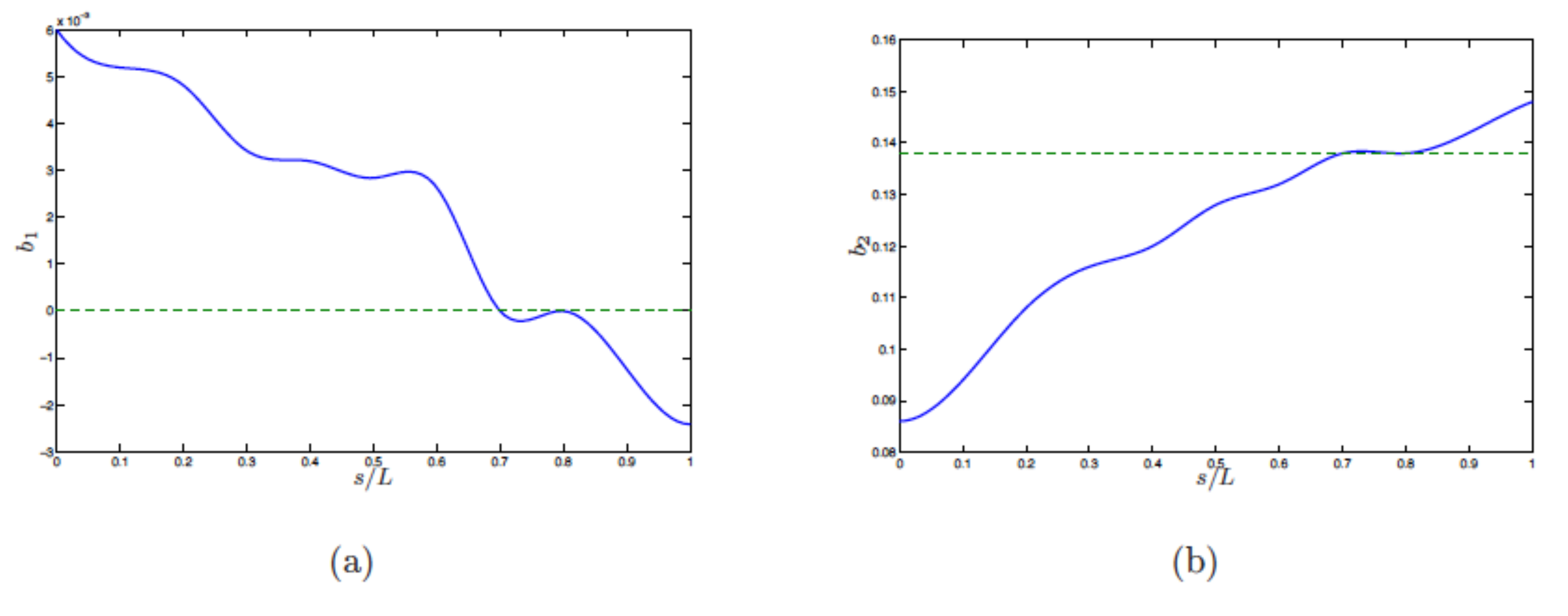}
\end{center}
\caption{Comparison of the terms (a) $b_1$ and (b) $b_2$,
  cf.~\eqref{eq:ab}, evaluated based on (solid line) the averaged
  solution of the time--dependent problem and (dashed line) the
  closure model described in Section \ref{sec:efs_closure},
  cf.~\eqref{eq:b2}. The data is shown as a function of the normalized
  arclength coordinate $s/L$ along the effective surface $\tg$
  measured from the top, and the values of the parameters are
  $V_d=1.0$, $r=0.25$ and $T=3$.}
\label{fig:b}
\end{figure}

Finally, in Figure \ref{fig:EF}d, we perform a comparison between the
effective free surfaces constructed using the algebraic closure model
from Section \ref{sec:efs_closure} and using the time--averaged
solutions of the original unsteady problem to evaluate the terms $a$
and $\b$ via relations \eqref{eq:AB} and \eqref{eq:ab}. The parameters
used in this case are $V_d = 1.0$, $r = 0.25$ and $T = 3.0$, and the
corresponding data for $\b$ obtained by averaging over 200 periods of
droplet impingement in the time--dependent case is shown in Figure
\ref{fig:b} (the data for the term $a$ is not shown as it vanishes by
construction in both cases, cf.~\eqref{eq:a2}). For comparison, in
Figure \ref{fig:b} we also indicate the values of the components of
$\b$ obtained from expressions \eqref{eq:AB}, and we see that the
predictions of the simple algebraic closure model developed in Section
\ref{sec:efs_closure} are not too far off from the actual data: as
regards the vertical component $b_2$, they are within the same order
of magnitude (Figure \ref{fig:b}b), whereas for the horizontal
component $b_1$ the difference is $\O(10^{-3})$ (Figure \ref{fig:b}a).
We add that for both $b_1$ and $b_2$ there exists a part of the
effective surface $\tg$, located towards the bottom of the liquid
column $\tilde{\Omega}_L$, where the agreement between the closure
model and the actual data is particularly good.  In Figure
\ref{fig:EF}d we note an overall fairly good agreement between the
effective surfaces obtained with terms $a$ and $\b$ evaluated in the
two different ways. This is, in particular, the case as regards the
top part of the liquid column $\tilde{\Omega}_L$ which, somewhat
interestingly, does not coincide with the region where the closure
model is the most accurate according to the data from Figure
\ref{fig:b}. We attribute this effect to the nonlinear and nonlocal
character of the averaged free--boundary problem \eqref{nsR2}.
Finally, we conclude that, despite its simplicity, our closure model
performs relatively well in the present problem.

\section{Conclusions}
\label{sec:final}

In this investigation we revisited the concept of ``effective free
surfaces'' arising in the solution of time--averaged hydrodynamic
equations in the presence of free boundaries
\cite{hw00,bp01a,bp01b,wo11}.  The novelty of our work consists in
formulating the problem such that there is a {\em sharp} interface
separating the two phases in the time--averaged sense, an approach
which appears preferable from the point of view of a number of
possible applications. The resulting system of equations is of the
free--boundary type and we also propose a flexible and efficient
numerical method for the solution of this problem which is based on
the shape--optimization formulation. Subject to some clearly stated
assumptions, the terms representing the average effect of the boundary
fluctuations appear in the form of interface boundary conditions, and
a simple algebraic model is proposed to close these terms (this is to
be contrasted with the ``classical'' Reynolds stresses which are
defined in the bulk of the flow).

This work is motivated by applications of optimization and optimal
control theory to problems involving free surfaces. In such problems
dealing with time--dependent governing equations leads to technical
difficulties, many of which are mitigated when methods of optimization
are applied to a {\em steady} problem with effective free surfaces.
The model problem considered in this study concerns impingement of
free--falling droplets on a liquid with a free surface in two
dimensions and is motivated by optimization of the mass and momentum
transfer phenomena in certain advanced welding processes, see Volkov
\emph{et al.}\cite{vplg09a}. The computational results shown in this
paper are, to the best of our knowledge, the first ever presented for
a problem of this type where the effective boundary has the form of a
sharp interface. The computed effective free surfaces exhibit a
consistent dependence on the problem parameters introduced via the
closure model, and despite the admitted simplicity of this model,
these results match well the effective surfaces obtained using data
from the solution of the time--dependent problem.

A key element of the proposed approach is a closure model for the
fluctuation terms representing the motion of the free surfaces.  The
model we developed here is a very elementary one resulting in simple
algebraic relationships. As in the traditional turbulence
research\cite{pope}, more advanced and more general closure models can
be derived based on the PDEs describing the transport of various
relevant quantities such as the turbulent kinetic energy, the
turbulent length scale, etc. In fact, such approaches have already
been explored in the context of free--boundary
problems\cite{bp01b,b02} leading to more general, albeit less
explicit, closure models than the model considered here.

In addition to investigating such more advanced closure models, our
future work will focus on quantifying the effect of and, ultimately,
weakening the assumptions employed to derive the present approach, so
that it can be applied to a broader class of problems, especially
interfacial. At the same time, we will seek to incorporate this
approach into the optimization--oriented models of complex
thermo--fluid phenomena occurring in welding processes, such as
discussed in Volkov \emph{et.~al}\cite{vplg09a}. While the present
investigation responded to needs arising from a certain class of
applications, it has also highlighted a number of more fundamental
research questions which it will be worthwhile to explore based on
even simpler model problems such as, e.g., capillary or gravity waves
on a flat interface.

\section*{Acknowledgments}

The authors wish to acknowledge generous funding provided for this
research by the Natural Sciences and Engineering Research Council of
Canada (Collaborative Research and Development Program) and General
Motors of Canada. The authors are also thankful to Dr.~J\'er\^ome
Hoepffner for interesting discussions, and to the two anonymous
referees for insightful and constructive comments.

\clearpage
\appendix

\section{Characterization of Cost Functional Gradients in 
Effective Surface Calculations}
\label{sec:efs_grad}

In this Appendix we obtain expressions for the gradient
$\bnabla_{\tg}\J$ of the cost functional \eqref{J} with respect
to the position of the effective interface $\tilde{\Gamma}_{LG}$.
Characterization of this gradient requires one to differentiate
solutions of governing PDEs system \eqref{nsR2} with respect to the
shape of the domain on which these solutions are defined. This is done
properly using tools of the shape--differential calculus 
\cite{sz92,dz01a} which are briefly introduced below. In the 
calculations we will assume that the problem parameters $\{V_d, r,
T\}$ are given. Hereafter primes ($'$) will denote perturbations
(shape differentials) of the different variables which is consistent
with the convention used in the literature\cite{dz01a}. Since
fluctuation variables do not appear in this Appendix, there is no risk
of confusion resulting from this abuse of notation.

\subsection{Shape Calculus}
\label{sec:efs_shape}

In the shape calculus perturbations of the interface geometry can be
represented as
\begin{equation}\label{x_eta}
\x(\eta,\x')=\x+\eta \x' \quad \text{for} \quad \x \in \tg(0),
\end{equation}
where $\eta$ is a real parameter, $\tg(0)$ is the original unperturbed
boundary and $\x' : \Omega \to \mathbb{R}^2$ is a ``velocity'' field
characterizing the perturbation. The G\^ateaux shape differential of a
functional such as \eqref{chi} with respect to the shape of the
interface $\tg$ and computed in the direction of perturbation field
$\x'$ is given by
\begin{equation}\label{J'}
\mathcal{J}'(\tg; \x')= \lim_{\eta \to 0}
\frac{\mathcal{J}(\x_{\tg}+\eta \x')-\mathcal{J}(\x_{\tg})}{\eta}.
\end{equation}
In the sequel we will need the following fundamental result concerning
shape--differentiation of functionals defined on smooth domains
$\Omega(\eta , \x')$ and on their boundaries, and involving smooth
functions $f$ and $g$ as integrands \cite{dz01a}
\begin{equation}
\begin{aligned}
\Bigg( \int_{\Omega (\eta, \x')} f \,d\Omega +\int_{\partial \Omega
  (\eta, \x')} g \,d\sigma \Bigg)'= & \int_{\Omega(0)} f' \,d\Omega
+\int_{\partial \Omega (0)} g' \,d\sigma  + \\
& \int_{\partial \Omega (0)} \left( f + \kappa\, g + \frac{\partial g}{\partial n}\right) (\x'\cdot\n)\, d\sigma,
\end{aligned}
\label{def_shape}
\end{equation}
where $f'$ and $g'$ are the shape derivatives of $f$ and $g$, and
$\kappa$ is the curvature of the boundary $\partial\Omega(0)$.

\subsection{Differential of Cost Functional}
\label{sec:efs_dJ}

In order to obtain the gradients $\nabla_{\tg}\mathcal{J}$ of the cost
functional \eqref{J} with respect to the control variable $\tg$, we
first need to obtain the G\^ateaux (directional) derivative of
$\J(\tg)$.  Using relation \eqref{def_shape} and substituting $f
\equiv 0$ and $g=\frac{1}{2}\chi^2$, we obtain the following
expression for the cost functional differential
\begin{equation}\label{cost_functional_diff}
\mathcal{J}'(\Gamma; \x')= \int_{\Gamma} \chi\,\chi' \,d\sigma + 
\int_{\tg} \left(
\frac{1}{2} \kappa \,\chi^2+\chi \,\frac{\partial \chi}{\partial n}\right) 
(\x'\cdot\n) \, d\sigma,
\end{equation}
where
\begin{multline}\label{chi'}
\chi'=\n'\cdot\asigL \cdot \n+\n\cdot \asigL' \cdot \n + \n\cdot\asigL \cdot \n'-
\n'\cdot\asigG \cdot \n  - \n \cdot\asigG' \cdot \n  - \n\cdot\asigG \cdot \n' \\ - \kappa'\,\gamma  
- 2 (\rho_L- \rho_G)\left[ \frac{r}{T^2} \sqrt{4 V_d^2T^2-\pi^2 r^2}- \frac{r^2}{V_d^2 T^4}\left( 4 V_d^2T^2-\pi^2 r^2\right) \right]  (\mathbf{e_y} \cdot \n) (\mathbf{e_y} \cdot \n').
\end{multline}
Using the following identities of shape calculus \cite{dz01a}
$\n'=-\bnabla_{\Gamma}(\x'\cdot\n)$ and $\kappa'=-\Delta_{\Gamma}
(\x'\cdot\n)$, where $\bnabla_{\Gamma}$ and $\Delta_{\Gamma}$ are the
tangential gradient and the Laplace--Beltrami operator, we obtain
\begin{multline}\label{J'_shape}
\mathcal{J}'(\Gamma;\x')= \int_{\tg}\Bigg[ \chi\, 
\Bigg( \big[\n\cdot\sig'\cdot\n\big]_L^G  - 
2 \bnabla_{\Gamma} (\x' \cdot \n) \cdot \asigL \cdot  \n + 2 \bnabla_{\Gamma} (\x' \cdot \n) \cdot  \asigG \cdot \n
+ \gamma \Delta_{\Gamma} (\x' \cdot \n)\Bigg) \\
+ 2 (\rho_L- \rho_G)\left[\frac{r}{T^2} \sqrt{4 V_d^2T^2-\pi^2 r^2}- \frac{r^2}{V_d^2 T^4}\left(  4 V_d^2T^2-\pi^2 r^2 \right) \right]   (\mathbf{e_y} \cdot \n) \left(\mathbf{e_y} \cdot   \bnabla_{\Gamma} (\x' \cdot \n) \right) \\ +
\left( \frac{1}{2} \kappa\, \chi^2+
\chi \,\frac{\partial \chi}{\partial n}\right) (\x' \cdot \n)\Bigg]
d\sigma.
\end{multline}
Considering G\^ateaux differential \eqref{J'_shape} and invoking the
Riesz representation theorem \cite{b77} allows us to extract the cost
functional gradient $\nabla \mathcal{J}(\tg)$ through the
following identity
\begin{equation}\label{riesz}
\mathcal{J}'(\Gamma; \x')= \Big \langle \nabla  \mathcal{J}(\tg), \zeta' \Big\rangle_{L_2(\tg)} 
= \int_{\tg} \nabla \mathcal{J}\, \zeta' \, d\sigma,
\end{equation} 
where for simplicity the $L_2$ inner product was used and $\zeta' =
(\x'\cdot\n)$ which implies that the gradient $\nabla \J$ is a
scalar--valued function describing the sensitivity to shape
perturbations in the normal direction. We note that expression
\eqref{J'_shape} contains terms which are already in the Riesz form
with the perturbation $(\x'\cdot\n)$ appearing as factor, in addition
to terms involving the shape derivatives of the state variables,
namely $\avp'$ and $\app'$. The presence of these terms makes it
impossible at this stage to use \eqref{J'_shape} to identify the
gradient $\nabla \mathcal{J}(\tg)$. In order to transform the
remaining part of relation \eqref{J'_shape} into a form consistent
with the Riesz representation \eqref{riesz} it is necessary to define
suitable {\em adjoint variables} and the corresponding adjoint system.

\subsection{Adjoint Equations}
\label{sec:efs_adj}
Consider the weak form of system \eqref{nsR2} for the variables
$\avp, \avp^* \in \mathbf{H}^1$ and $\langle p \rangle , \langle p \rangle ^* \in
\text{L}_2$
\begin{eqnarray}\nonumber
\int_{\Omega_L}\left [\rho_L \avp \cdot \bnabla \avp- \bnabla \cdot \asigLmu -
\rho_L \g\right ] \cdot \avp^{*}- ( \bnabla \cdot \avp) \langle p \rangle ^{*} d\x
+ \\ \label{ns_weak}
\int_{\Omega_G}\left[\rho_G\avp\cdot \bnabla \avp- \bnabla \cdot \asigGmu -
\rho_G \g\right] \cdot \avp^{*}- ( \bnabla\cdot\avp) \langle p \rangle ^{*} d\x=0
\end{eqnarray}
After integrating the second--order terms by parts, \eqref{ns_weak} becomes
\begin{multline}
\int_{\Omega_L}\left[\rho_L \avp\cdot \bnabla \avp- 
\rho_L \g\right] \cdot \avp^{*}- \avp \cdot \bnabla \langle p\rangle^*  +
\asigLmu: \bnabla \avp^{*} d\x
+ 
\int_{\Omega_G}\left[\rho_G\avp\cdot \bnabla \avp- \rho_G \g\right] \cdot \avp^{*}
\\ \label{ns_weak_int} 
- \avp\cdot\bnabla \langle p \rangle^*  +  \asigGmu: \bnabla \avp^* \, d\x - 
 \int_{\tg} \n\cdot\asigLmu\cdot \avp^{*} d\sigma -  \int_{\tg} \n\cdot\asigGmu \cdot \avp^{*} d\sigma \\
 - \int_{\tg} \n\cdot \avp \, \langle p \rangle^* d\sigma =0
 \end{multline}
Next, using relation \eqref{def_shape} and shape differentiating 
\eqref{ns_weak_int}, we obtain
\begin{gather}
\int_{\Omega_L} \left[\rho_L \avp \cdot\bnabla
\avp' +\rho_L \avp'\cdot \bnabla
\avp\right]\cdot \avp^{*}-  \bnabla \langle p \rangle ^* \cdot\avp' +
\asigLmu': \bnabla \avp^* d\x + \\ \nonumber
\int_{\Omega_G} \left[\rho_G \avp \cdot\bnabla
\avp' +\rho_G \avp'\cdot \bnabla \avp\right]
\cdot \avp^{*}-  \bnabla \langle p \rangle ^*\cdot\avp' +
\asigGmu': \bnabla \avp^* d\x+ \mathcal{I}=0 
\label{ns_weak_int_a}
\end{gather}
where 
\begin{multline}\label{I}
\mathcal{I}\triangleq \int_{\tg} \Bigg\{ \left[\rho_L \avp\cdot \bnabla \avp- 
\rho_L \g\right] \cdot \avp^{*}- \avp \cdot  \bnabla \langle p \rangle^* +
\asigLmu: \bnabla \avp^*+\kappa\, \n\cdot\asigLmu\cdot\avp^* \\ + 
\frac{\partial
}{\partial n} \left( \n\cdot\asigLmu\cdot\avp^*\right)  
 +\left [\rho_G \avp\cdot \bnabla \avp- 
\rho_G \g\right] \cdot \avp^{*}- \avp\cdot\bnabla \langle p \rangle^* +
\asigGmu: \bnabla \mathbf{v^*}+
\kappa\, \n\cdot\asigGmu\cdot\avp^* \\ + 
 \frac{\partial
}{\partial n} \left( \n\cdot\asigGmu\cdot\avp^* \right) \Bigg \} ( \x' \cdot \n)\, d\sigma
- \int_{\tg} \Big [\n\cdot\asigLmu'\cdot\avp^*+
\n'\cdot\asigLmu\cdot\avp^*+\n\cdot\asigGmu'\cdot \avp^*+ 
\n'\cdot\asigGmu\cdot\avp^* \\
+\n\cdot \avp'\, \langle p \rangle^*+ \n'\cdot\avp \,\langle p \rangle ^* \Big]\, d\sigma.
\end{multline}
Performing one more time integration by parts in \eqref{ns_weak_int_a} we obtain
\begin{multline} \label{ns_weak_adjoint}
\int_{\Omega_L} \Bigg\{ \left[-\rho_L \avp\cdot\bnabla
\avp^* +\rho_L \avp^*\cdot (\bnabla \avp)^T\right]\cdot \avp'- 
\app' \, \bnabla\cdot\avp^*  - \mu\, \avp' \cdot \Delta \avp^* - 
\avp' \cdot \bnabla \app^* \Bigg\} d\x   \\ +
  \int_{\Omega_G} \Bigg\{\left[-\rho_G \avp\cdot\bnabla
\avp^* +\rho_G \avp^*\cdot (\bnabla \avp)^T\right]\cdot \avp'- 
\langle p \rangle'\, \bnabla\cdot\avp^* \, - \mu\,\avp'\cdot \Delta \avp^* \\ 
-  \avp' \cdot  \bnabla \app^*\Bigg\}  d\x +
\mu \int_{\tg}
\left[(\n \cdot \bnabla \avp^*)\cdot \avp'+(\n\cdot (\bnabla \avp^*)^T)\cdot \avp' \right]d\sigma \\
+ \mu \int_{\tg}
\left[(\n \cdot \bnabla \avp^*)\cdot \avp' +
\n\cdot (\bnabla \avp^*)^T\cdot \avp'\right] d\sigma+ \mathcal{I} = 0,
\end{multline}
where $\avp^*$ and $\langle p \rangle ^*$ can be identified as the adjoint
variables with respect to $\avp$ and $\langle p \rangle $, provided they satisfy
the following adjoint equations
\begin{subequations}\label{ns_adjoint_liquid}
\begin{alignat}{2}
\rho_L \avp\cdot\bnabla \avp^* &=\rho_L
\avp^*\cdot (\bnabla \avp)^T - ( \bnabla \langle p \rangle ^*)- \mu
\Delta \avp^* \quad && \text{in} \  \Omega_L, \label{adjoint_continuity_liquid}\\
\bnabla\cdot \avp^* &=0  \quad && \text{in} \  \Omega_L, \label{adjoint_momentum_liquid}
\end{alignat}
\end{subequations}
\begin{subequations}\label{ns_adjoint_gas}
\begin{alignat}{2}
\rho_G \avp\cdot\bnabla
\avp^* &=\rho_G \avp^*\cdot (\bnabla \avp)^T -( \bnabla \app^*)- \mu
\Delta \avp^*\quad &&
\text{in} \  \Omega_G, \label{adjoint_continuity_gas}\\
\bnabla\cdot  \avp^* &=0  \quad &&
\text{in} \  \Omega_G. \label{adjoint_momentum_gas}
\end{alignat}
\end{subequations}
Substituting for $\n'$ in \eqref{I} we can simplify the
expression for $\mathcal{I}$ as follows
\begin{multline}\label{I_a}
\mathcal{I} = \int_{\tg}\Bigg\{\left[\rho_L \avp\cdot \bnabla \avp- 
\rho_L \g\right] \cdot \avp^{*}- \avp\cdot\bnabla \app^*  +
\asigLmu: \bnabla \avp^*+ \
\kappa\, \n\cdot\asigLmu\cdot\avp^* \\
+ \frac{\partial
}{\partial n} \left( \n\cdot\asigLmu\cdot\avp^*\right)  
+\left[\rho_G \avp\cdot \bnabla \avp- 
\rho_G \g\right] \cdot \avp^{*}- \avp\cdot\bnabla \app^*  +
\asigGmu: \bnabla \avp^* \\+ \kappa\, \n\cdot\asigGmu\cdot\avp^*+ 
 \frac{\partial
}{\partial n} ( \n\cdot\asigGmu\cdot\avp^*)\Bigg\}( \x' \cdot \n)\, d\sigma
-  \int_{\tg}
\Bigg[\left[\n\cdot\asigmu'\cdot\avp^*\right]_L^G-
\bnabla_{\Gamma} (\x'\cdot \n)\cdot\asigLmu\cdot\avp^* \\ -
\bnabla_{\Gamma} (\x'\cdot \n)\cdot \asigGmu\cdot\avp^* - 
\bnabla_{\Gamma} (\x'\cdot \n)\cdot \avp \, \app^*  \Bigg]\, d\sigma .
\end{multline}
Imposing the boundary conditions
\begin{subequations}\label{bc_adjoint}
\begin{align}
\avp^*\Big|_L&=\avp^* \Big|_G= - \chi\, \n, \label{bc_adjoint_a}\\
\avp^* &=0, \label{bc_adjoint_b}
\end{align}
\end{subequations}
the expression for the differential of the cost functional becomes
\begin{multline}\label{gradient}
\mathcal{J}'(\Gamma, \x')= \int_{\tg}\Bigg\{- \bnabla_{\Gamma} (\x'\cdot \n)\cdot \asigLmu\cdot \n  -
 \bnabla_{\Gamma} (\x'\cdot\n)\cdot \asigGmu\cdot \n 
 + \gamma \Delta_{\Gamma} (\x'\cdot\n)
 \\
 + 2 (\rho_L- \rho_G)\left[ \frac{r}{T^2}  \sqrt{4 V_d^2T^2-\pi^2 r^2} - \frac{r^2}{V_d^2 T^4} \left(4 V_d^2T^2-\pi^2 r^2 \right) \right] (\mathbf{e_y} \cdot \n) \left( \mathbf{e_y} \cdot  \bnabla_{\Gamma} (\x'\cdot \n) \right)
  \\  + \left(
\frac{1}{2} \kappa\, \chi^2+\chi \frac{\partial \chi}{\partial n}\right)
(\x'\cdot\n)
 -   \Bigg[\left[\rho_L \avp\cdot \bnabla \avp- 
\rho_L \g\right] \cdot \avp^{*}-   \avp\cdot \bnabla \app^*  +
\asigLmu: \bnabla \mathbf{v^*} \\ 
+  \kappa\, \n\cdot\asigLmu\cdot \avp^*+  \frac{\partial
}{\partial n} ( \n\cdot\asigLmu\cdot\avp^*)  
+ \left[\rho_G \avp\cdot \bnabla \avp- 
\rho_G \g\right] \cdot \avp^{*}- \avp\cdot  \bnabla \app^*  +
\asigGmu: \bnabla \avp^*
\\ 
+ \kappa  \, \n \cdot \asigGmu\cdot\avp^*+\frac{\partial
}{\partial n} \left( \n\cdot\asigGmu\cdot\avp^*\right)\Bigg] ( \x' \cdot \n)  \Bigg\}  d \sigma
\end{multline}
which is now consistent with Riesz's representation \eqref{riesz}.
Finally, after applying tangential Green's formula\cite{dz01a} to the terms
involving $\bnabla_{\Gamma} (\x'\cdot\n)$, the cost functional
gradient can be identified as follows
\begin{multline}\label{final_gradient}
\nabla\mathcal{J}(\tg)=\Bigg[ - \n\cdot \bnabla_{\Gamma}\cdot\asigLmu-
\n\cdot\bnabla_{\Gamma}\cdot\asigGmu + \left(
\frac{1}{2} \kappa \,\chi^2+\chi\, \frac{\partial \chi}{\partial n}\right) \\
+ 2 (\rho_L- \rho_G)\left[ \frac{r}{T^2} \sqrt{4 V_d^2T^2-\pi^2 r^2} - \frac{r^2}{V_d^2 T^4} \left( 4 V_d^2T^2-\pi^2 r^2 \right) \right]  ( \mathbf{e_y} \cdot \n )
-  \left[\rho_L \avp\cdot \bnabla \avp- 
\rho_L \g\right] \cdot \avp^{*}  \\ -   \avp\cdot \bnabla \app^*  +
\asigLmu: \bnabla \avp^* +\kappa\, \n\cdot\asigLmu\cdot\avp^*+ 
\frac{\partial
}{\partial n} \left( \n\cdot\asigLmu\cdot\avp^*\right)  
+\left[\rho_G \avp\cdot \bnabla \avp- 
\rho_G \g\right] \cdot \avp^{*} \\ -\avp\cdot \bnabla \app^* +
\asigGmu: \bnabla \avp^*+
\kappa \,\n\cdot\asigGmu\cdot\avp^*+\frac{\partial
}{\partial n} \left( \n\cdot\asigGmu\cdot\avp^* \right)+
\gamma  \kappa   \Bigg] \quad \text{on} \quad
\tg\cdot
\end{multline}
Consistency of this expression for the cost functional gradient is
demonstrated in Section \ref{sec:efs_kappa}.

\end{document}